\documentclass[printer]{aa} 
\usepackage{graphicx}
\usepackage{txfonts}
\usepackage{longtable}
\usepackage{natbib}
\usepackage[a4paper,breaklinks]{hyperref}

\begin{document}
   \title{Extracting interstellar diffuse absorption bands from cool star spectra}

   \subtitle{Application to bulge clump giants in Baade's window}

   \author{
   	H.-C. Chen\inst{1}, 
          R. Lallement\inst{1},
          C. Babusiaux\inst{1},
          L. Puspitarini\inst{1},
          P. Bonifacio\inst{1},
          \and
          V. Hill\inst{2}
          }

   \institute{GEPI Observatoire de Paris, CNRS, Universit\'e Paris-Diderot, 
              Place Jules Janssen, F-92195 Meudon Cedex\\
              \email{hui-chen.chen@obspm.fr}
         \and
Universit\'e de Nice Sophia Antipolis, CNRS,
Observatoire de la C\^ote d'Azur, Laboratoire Cassiop\'ee, B.P. 4229, 
06304 Nice Cedex 4, France
             }

   \date{Received 2012; accepted}

% \abstract{}{}{}{}{} 
% 5 {} token are mandatory
 
  \abstract
   {}
   { Interstellar (IS) absorption lines or diffuse bands are usually extracted from early-type star spectra because they are characterized by smooth continua. However this introduces a strong limitation on the number of available targets, and reduces potential studies of the IS matter and the use of absorptions for cloud mapping. 
    }
   {We have developed a new and automated fitting method appropriate to interstellar absorptions in spectra of cool stars that possess stellar atmospheric parameters. We applied this method to the extraction of three diffuse interstellar bands (DIBs) in high resolution VLT FLAMES/GIRAFFE spectra of red-clump stars from the bulge. 
By combining  all  stellar synthetic spectra, 
HITRAN-LBLRTM atmospheric transmission spectra and diffuse 
band empirical absorption profiles, we determine the  
6196.0, 6204.5, and 6283.8 \AA\  DIB strength toward the 219 target stars and discuss the sources of uncertainties. 
In order to test the sensitivity of the DIB extraction, we inter-compare the three results and compare the DIB equivalent widths with the reddening derived from an independent extinction map based on OGLE  photometric data. We also derive the three independent color excess E(B-V) estimates based on the DIB strengths and average linear correlation coefficients previously established in the solar neighborhood and compare with the OGLE photometric results.}
   {Most stellar spectra could be well reproduced by the composite stellar, atmospheric
   and interstellar models. Measurement uncertainties on the EWs are smaller for the broad and strong 6283.8 \AA\ DIB, and are of the order of 10-15 \%. Uncertainties on the two narrow and weaker DIBs are larger, as expected, and found to be highly variable from one target to the other. They strongly depend on the star's radial velocity.  
DIB-DIB correlations among the three bands demonstrate that a meaningful signal is extracted. For the 6284 and  6204 \AA\ DIBs, the star-to-star variability of the equivalent width (EW) also reflects features of the OGLE extinction map.
The three independent extinction estimates deduced from the EWs and solar neighborhood correlation coefficients agree with each other within 20 \% only, which probably reflects 
the fact that they belong to different {\it families}. 
The estimated average color excess is also 
compatible with the photometric determination.
}
 {This work demonstrates the feasibility of the method of ISM DIB extraction in cool star spectra, based on synthetic spectra.
It confirms that DIB measurements and local DIB-extinction calibrations can provide rough, first order estimates of the  towards distant targets. 
}
 

  \maketitle
%
%=============
\section{Introduction}

 DIBs and interstellar absorption lines have been almost exclusively extracted from early-type star spectra, using the fact that stellar lines are broad, shallow and in limited number, and do not contaminate the IS lines. Their continua can be easily fitted without any need for stellar models. However, this is a strong limiting factor since it reduces considerably the number of potential targets and the information that can be obtained on the source regions of the absorptions and especially the spatial resolution that can be achieved. In the case of the DIBs, it is limiting studies of the DIB response to the radiation field and to the local physical properties within the IS clouds, while such studies are promising tools in the search for the DIB carriers (e.g. Vos et al. , 2011).
Moreover, with the advent of stellar spectroscopic surveys at increasing resolution, the use of multi-object spectrographs and the perspective of forthcoming Gaia parallaxes, a large amount  of line-of-sight integrated absorption data will become available. Extracting absorption characteristics towards all targets, including cool stars, would allow a better mapping of the galactic ISM, because it would considerably increase the achievable spatial resolution. For such mapping, interstellar lines of atoms, simple molecules, and DIBs are available, since both contain information on the amount of IS matter. While the former are well identified, carriers of the latter remain unknown, despite their early discovery 
\citep[e.g.]{herbig95,JD94,salama96,fulara00,snow11,fri2011,hobbs2008,hobbs2009}. 
Still, this lack of carrier identification and the fact that most of them are only weakly correlated with extinction or gas column are not incompatible with their use as tracers of IS clouds in the same way IS lines are used, i.e. the radial gradients of their strengths allow to locate IS clouds in distance and build 3D maps. In some respects, DIBs may be advantageous ISM tracers: they are numerous, thus observable at various instrumental settings, they are often broad, and in general they are unsaturated. Recent data analyses suggest that the average ratios between the DIBS and the extinction do not seem to vary significantly within the first kiloparsec \citep{fri2011,Vos11}. Finally, making use of cooler stars to extract their strengths reduces the dispersion around the average relationship, because extreme ionizing field effects linked to bright and hot targets are avoided \citep{rai2012}. For this reason, extracting DIB information from cool stars would be another improvement as far as mapping is concerned. 

Here we present a new method of DIB measurement based on cool star synthetic spectra, and as a test case we apply this technique to three different diffuse bands and 220 targets located in the galactic bulge that have the advantage of possessing a precise determinations of their atmospheric parameters. We selected the 6196.0, 6204.5, and 6283.8 ${\AA}$ DIBs that are the three strongest absorptions contained in the observed spectral interval. Those DIBs have been already widely studied (see e.g. the works of  Galazutdinov et al 2002, Cami et al, 1997, Friedman et al, 2011, Vos et al, 2011). 
We estimate the statistical and systematic uncertainties on the DIB equivalent widths (EWs) and discuss potential improvements. We compare the measured EW angular pattern over the field with an independent mapping of the reddening based on photometric data of stars close to the galactic center (GC). We finally use average DIB-reddening empirical relationships previously established  in the Sun vicinity to derive three independent estimates of the reddening towards all targets, inter-compare the three  estimates and finally compare with the photometric determination.

This article is organized as follows. Data are presented in section 2. In section 3 we describe the new method of  interstellar absorption band extraction and its ingredients. 
In Section 4, we show  the model adjustments for selected target stars and discuss the results. 
In Section 5, we discuss the relations between the DIBs and between the DIBs and the extinction obtained by photometry. Section 6 presents color excess estimates based on the DIBs and nearby star studies.
Finally, we summarize the results and discuss the perspectives in the last section.
%=============

\section{Data}
We used the sample of 219
bulge red clump giants of \citet{hil2011}. Those stars have been observed 
with FLAMES/GIRAFFE at the VLT and are located within a 12 arcmin radius field in Baade's Window  (l=0.8, b=-4). We used here the spectra obtained with the GIRAFFE HR13 setup, leading to a resolution of R=22500 and a spectral coverage spanning from 6120 to 6405 $\AA$, that contains three well known diffuse bands of this study. The target selection, observations, data reduction and stellar parameters determination are described by \cite{hil2011}. Those stellar parameters were determined by these authors from photometric and spectroscopic data.  Here we use the effective temperature, gravity, micro-turbulence and metallicity resulting from their analysis, whose full ranges for the present sample are listed in Table~\ref{stparameter}. The different exposures were observed within a week interval so that we can neglect at this resolution variations of the radial velocity difference between the telluric and stellar lines and work with the final co-added spectra. Signal-to-noise ratios vary between $\simeq$ 30 and 77 per pixel (the Giraffe pixel-size is 0.07 \AA\ ). 

\begin{figure}
   \centering
   \includegraphics[width=9cm]{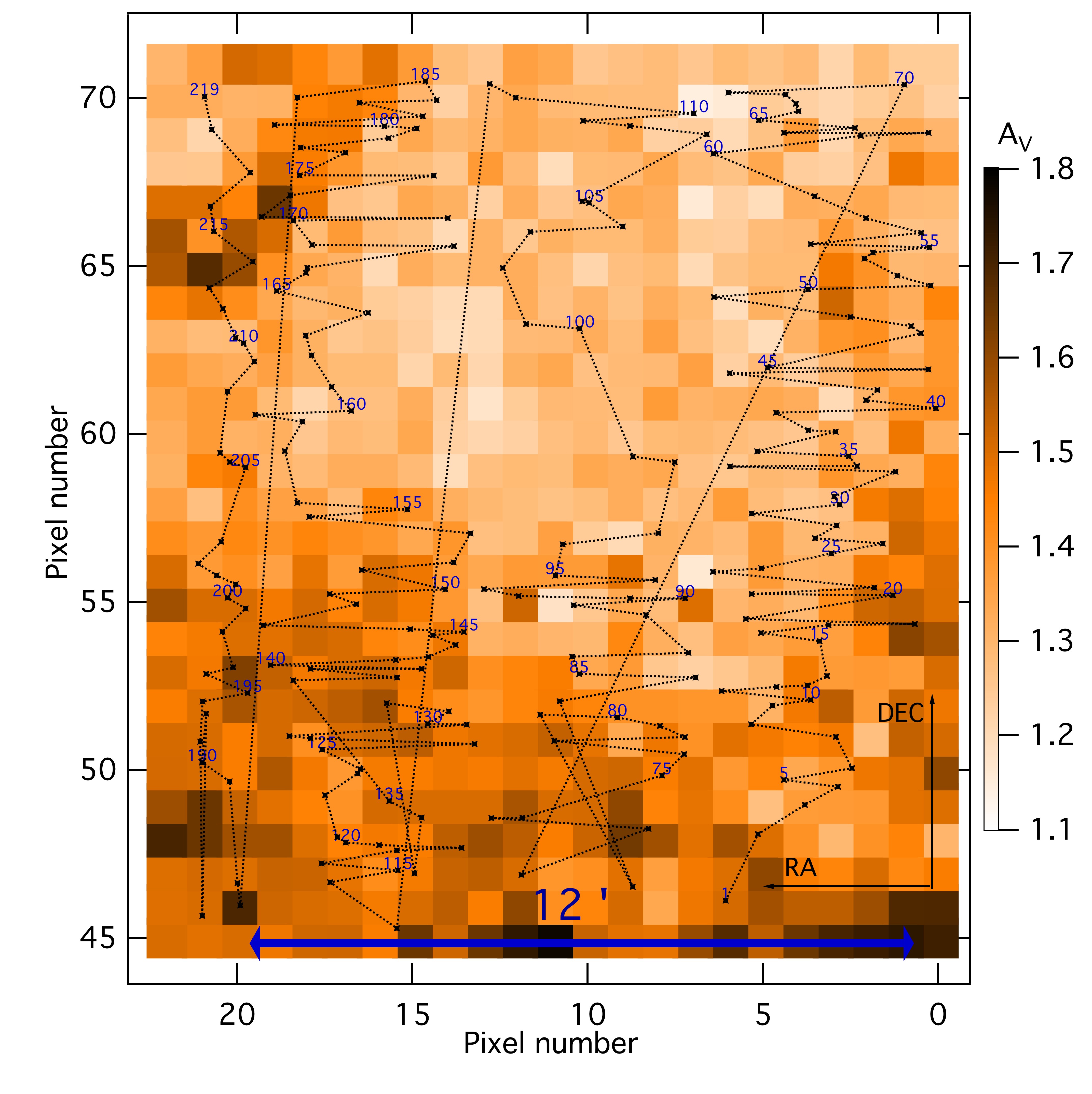}
     \caption{Target stars used in this study (black crosses) superimposed on the extinction map from Sumi (2004). The map is centered on (l,b)=(0.8,-4). Each pixel is 0.6'x0.6'. The extinction scale is displayed in fig \ref{Layout2gray}. Stars are numbered according to Table \ref{DIB_index}. Dotted lines link stars in increasing order of their ID number. Only one out of five numbers is indicated for clarity.}
         \label{sumi_map}
   \end{figure}

This region benefits from a high resolution differential extinction map that has been derived from the OGLE-II red clump giants photometry by \cite{sum2004}. 
Assuming a mean reddening corrected colour for the red clump, \cite{sum2004} divided each field in small sub fields, computed the mean observed colour of the red clump in each sub field and derived the reddening E(V-I). 
The resulting extinction map has a spatial resolution of 33 arcsec in our field. For comparison purposes we have interpolated within the \cite{sum2004} reddening grid to infer its value at each target location. Note that \cite{sum2004} indicates a potential zero-point offset for Av that may amount to 0.05. The distribution of the target stars is shown superimposed on the \cite{sum2004} extinction map
in Fig. \ref{sumi_map}.

\begin{table}
\caption{Stellar parameters. T$_{eff}$ is the effective temperature. 
	log $g$ is the gravity. $v_t$ is the micro-turbulent velocity. $[Fe/H]$ is the metallicity.}
\label{stparameter} 
\centering                          
\begin{tabular}{c c c c}  
%\hline
parameters & min & max & unit \\ 
\hline                        
   T$_{eff}$ & 4270 & 5448  & K\\
   log $g$ & 1.97 & 2.46 & \\
   v$_{t}$ & 0.8 & 1.8    &  km/s \\
   $[Fe/H]$ & -1.13 & 0.71  &  \\ 
\hline  
\end{tabular}
\end{table}

\section{Fitting procedure}

Our approach is to model and adjust simultaneously 
the spectrum of the background star, the DIB and the atmospheric
transmission  
 in one unique step. In the opposite case of hot stars, extracting the IS absorption was done in two or three steps. The correction for the telluric absorption, if necessary, was done first and independently of the IS absorption fitting. Synthetic transmission
spectra were adjusted through computation of the mean ratio between observed and modeled equivalent widths of specific lines (Lallement et al. 1993), or by minimizing the length 
(defined as the sum of flux differences between consecutive points)
of the residual spectrum 
obtained after dividing by the model atmospheric transmission (Raimond et al. 2012). None of those methods is necessary here, since the atmospheric transmission model is adjusted in velocity and airmass during the global adjustment. 
In a second step the smooth continuum of the hot star was fitted to a function, and in a last step the DIB strength was measured using the normalized spectrum.
Characterizing the DIB strength has been done in several ways, the two main ones being measures of the central depth and of the equivalent width. We refer to the detailed discussion of this point by Friedman et al (2011). We have chosen here the equivalent width, and, 
as we will see, this is the quantity that directly comes out from our cool star analysis.

In the case of cool stars, the presence of  numerous, deep and narrow 
stellar lines precludes the application of the simple continuum 
fitting procedure that is used for stars of 
earlier spectral types. 
Identifying and correcting the stellar lines one by one 
in the IS absorption region is not practical. 
Furthermore, if the stellar lines and the IS 
absorption do overlap, it is very difficult or even
impossible to separate them, hence our approach using a global adjustment. 
Our composite model here is the product of a 
synthetic stellar spectrum, a synthetic telluric 
transmission, and an empirical model for the DIB absorption. 
The model  allows for wavelength shifts 
between those three spectra to take into account the stellar radial 
velocity, the Earth motion, and the IS radial velocity. 
The giant stars used as background sources in the
present investigation do not have measurable 
projected rotational velocities at this resolution.
Sizeable projected rotational velocity in the background star
could be modelled as well, if necessary.
The combination of the three models is convolved by a Gaussian profile to take into account 
the instrumental spectral 
resolution and is adjusted to the data in the DIB region, 
with the DIB strength, the DIB center location, 
and an atmospheric  transmission scaling (see below) being the free parameters.

The synthetic stellar spectra, S($\lambda$),
were computed from an   $\textsf{ATLAS}$ 9 
model atmosphere using the  $\textsf{SYNTHE}$ suite   (Kurucz 2005).
We used the Linux port of all the codes (Sbordone et al. 2004; Sbordone 2005).
Atomic and molecular 
data have been taken
from the data base on  Kurucz's website\footnote{\url{http://kurucz.harvard.edu}}.
For each star we computed
a synthetic spectrum with 
the stellar parameters obtained by \cite{hil2011}, i.e.
the effective temperatures, gravity, metallicity, and 
microturbulence, summarized in Table~\ref{stparameter}. 
The stellar radial velocity is taken into account by doppler shifting the 
computed spectrum to the appropriate radial velocity.

The synthetic telluric transmissions are computed by means of the LBLRTM code (Line-By-Line Radiative Transfer Model, Clough et al 2005), using the molecular  database HITRAN (HIgh-Resolution TRANsmission molecular absorption, Rothman et al. 2009). Here we have used the transmission T$_{0}$($\lambda$) computed for a standard atmosphere and for the altitude of the observatory, and we have assumed that airmass variations from star to star simply result in a variable coefficient $\alpha$ for the transmission, with T($\lambda$) = T$_{0}$($\lambda$)$^{\alpha}$. The projected Earth motion is taken into account by 
doppler shifting the transmission.

Finally, the profiles of the 6196.0, 6204.5, and 6283.8 ${\AA}$ DIBs are derived from a high resolution (R=48,000), high signal to noise survey of early-type stars recorded with the FEROS spectrograph at the 2.2m ESO/Max Planck Institute Telescope in La Silla. Those profiles do not reveal the fine structure that is known to be present \cite{gala08} but this is not necessary here since we deal with spectra at a resolution that is lower than the one of FEROS.
The  shape of the 6283.8 ${\AA}$ DIB, which is contaminated by strong telluric molecular oxygen lines, has been derived by Raimond et al. (2102) by averaging a large number of individual shapes obtained after division by an adjusted telluric template. Such a use of a synthetic atmospheric model is allowed by the fact that this DIB is
 relatively strong and much broader than the telluric lines. In the same way, this DIB is here broader than the stellar lines and thus relatively easy to detect and measure (see figures \ref{noline}, \ref{oneline},\ref{weakDIB}). 
The two, much weaker DIBs 6196.0 and 6204.5 ${\AA}$ are in regions that are free of contaminating telluric lines. Their shapes have been derived by means of an automated fitting method appropriate for early-type stars (Puspitarini and Lallement, 2012). About ten individual profiles extracted from the best FEROS spectra have been averaged to produce those templates. 
Since those two DIBs are narrower than the 6283.8 ${\AA}$ DIB, here in the case of cool stars they are much more difficult to measure due to the overlap with the similarly narrow stellar lines. 

Neglecting emissions from the cloud in comparison with the stellar flux, DIB transmission profiles D($\lambda$) can be expressed as exp(-$\tau(\lambda))$, where $\tau(\lambda)$ is the optical thickness as a function of wavelength.
 $\tau(\lambda)$ is proportional to the column N of absorbers and can be expressed as $\tau(\lambda)=N / N_{0}  \tau_{0}(\lambda)$, $\tau_{0}(\lambda)$ and $N_{0}$ being some related optical thickness and column of reference. Calling $\beta$ the $N/N_{0}$ ratio,  a quantity  proportional to the column 
density of the DIB carrier, one has:
 
D($\lambda$)=D$_{0}(\lambda_{D})^{\beta}$, 
where D$_{0}$($\lambda$) = exp(-$\tau_{0}(\lambda))$ is the reference profile derived from 
the FEROS analyses, and $\lambda_{D}$ is the 
wavelength shifted by 
the radial velocity of the main absorbing medium. 
The DIB equivalent width  is by definition:
\begin{equation}
  W= \int \frac{I_{0}(\lambda)-I(\lambda)}{I_{0}(\lambda)} d\lambda = \int(1-exp(-\tau(\lambda)))  d\lambda 
\end{equation}
 where $I_{0}(\lambda)$ and $I(\lambda)$ are the unabsorbed and absorbed intensities. 
  Within the weak absorption regime appropriate here, $\tau$ is small,  $exp(-\tau) \simeq 1- \tau$ and  the equivalent width W is approximated by
  \begin{equation}
   W= \int \tau(\lambda) d\lambda = \int \beta \tau_{0}(\lambda) d\lambda = \beta W_{0}
\end{equation}
where $W_{0}$ is the equivalent width of the line of reference. 

We have chosen to list the results under the form of the product $\beta W_{0}$, which has the advantage of being both truly proportional to the absorbing column and to be an equivalent width, i.e. widely used, meaningful quantity. For the broader DIB, 6284 \AA\,  the relative difference between $\beta W_{0}$ and the actual EW is here smaller than 4.5 \% , a value reached for equivalent widths of the order of 1.2 \AA\ . For the narrower DIB, 6196 \AA\,  the departure from the linear regime for the EW occurs at a smaller optical thickness, however the DIB is weak, and finally  the relative difference between $\beta W_{0}$ and the actual EW is smaller than 4. \% , a value reached for equivalent widths of the order of 70 m\AA\ .

After adjustment of the coefficient $\beta$ through spectral fitting, the DIB equivalent width is thus simply computed as the product of the equivalent width of the reference profile $W_{0}$
by the coefficient $\beta$. 
We also assume here that most of the absorbing medium originates from clouds with a similar radial motion, or equivalently that the DIB is negligibly broadened by the line-of-sight velocity structure. This assumption is legitimated here by the fact that most of the absorption  arises within 1,500 pc from the Sun, as derived from the extinction model of Marshall et al. (2006) and that for those low l,b coordinates the projection of the gas motions onto the line-of-sight remains small with respect to the DIB spectral width. 

\begin{figure}
   \centering
   \includegraphics[width=9cm]{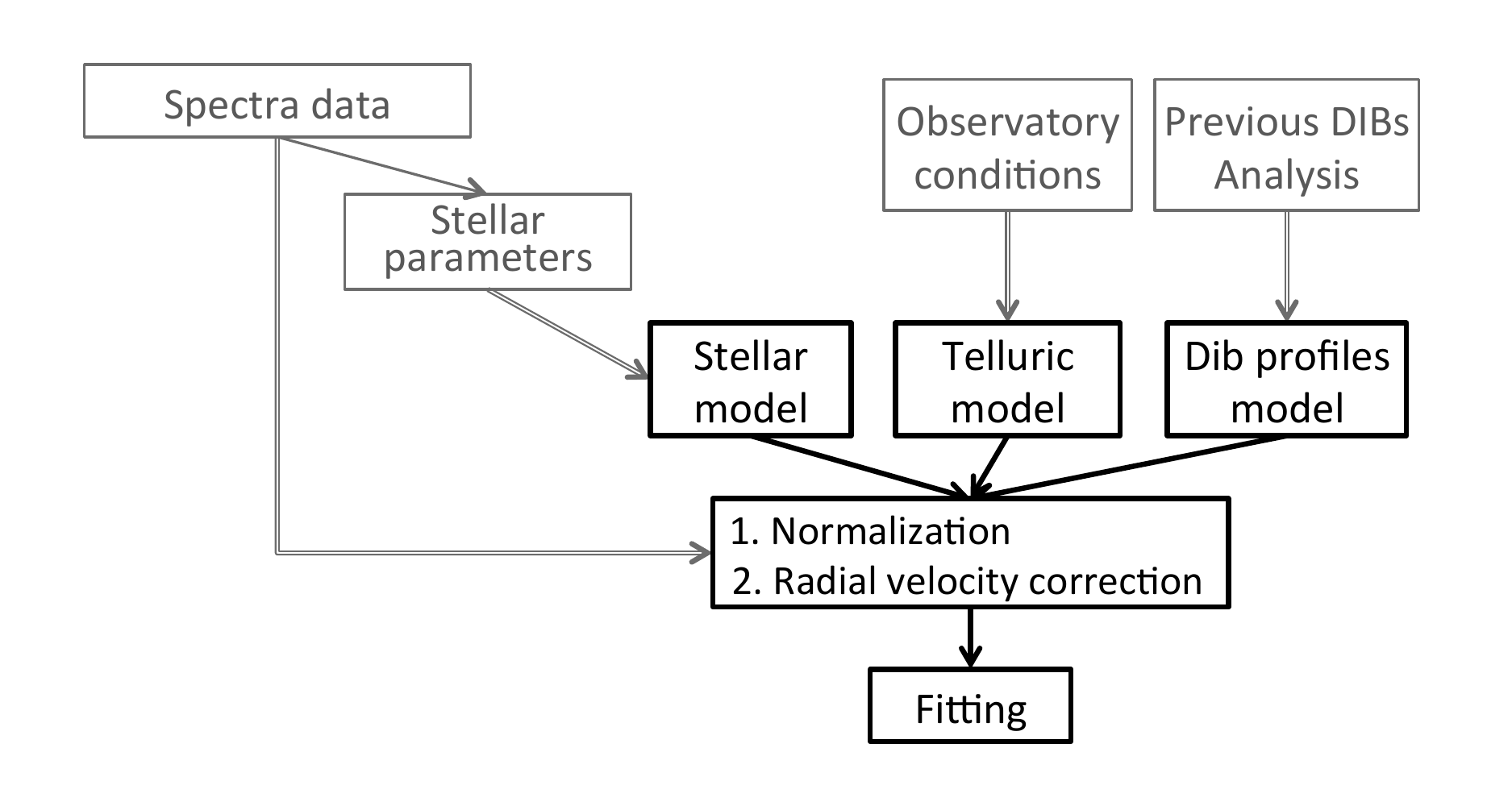}
     \caption{The procedure flowchart of the fitting.
              }
         \label{flowchart}
   \end{figure}

Figure \ref{flowchart} illustrates the procedure for this study. The stellar parameters are used as input for the stellar synthetic models and the stellar radial velocity is taken into account by an appropriate shift.  The atmospheric synthetic model of reference is calculated based on the position and altitude of the observatory. While doing the fitting, the program computes the product $\kappa$ S($\lambda_{S}$) T$_{0}(\lambda_{E})^{\alpha}$ 
D$_{0}(\lambda_{D})^{\beta}$, with $\kappa$ being a scaling factor,
$\lambda_S$ is the wavelength shifted by the stellar
radial velocity, $\lambda_E$ by the Earth's radial velocity and
$\lambda_D$ by the radial velocity of the DIB absorbing medium, 
and convolves the product of the three models by a Gaussian instrumental function appropriate for the resolution of GIRAFFE . 

Telluric absorptions should be identical for all fibers for the same exposure, and we also expect the DIB radial velocity to vary very little over the narrow field. Still, in order to test the adjustment and allow for small uncertainties in the three wavelength shifts, we started by allowing all parameters $\kappa$, $\alpha$, $\beta$, $\lambda_{S}$, $\lambda_{E}$ and $\lambda_{D}$ to freely vary, and performed a least mean squares adjustment to the data for each star. All the fittings were done in an automated way. The results revealed a large majority of identical or very close values for the  $\lambda_{E}$ shift (actually a null value for a fit in the earth frame) and for  $\alpha$. As mentioned previously this was expected for both $\alpha$, that responds to the airmass, and 
$\lambda_{E}$ that is a function of the date. We extracted those two parameters and kept them fixed for the second adjustment phase. We could also check from this first attempt that the values found by fitting for  $\lambda_{S}$ differ from the radial velocities from Hill et al. (2011) by very small velocity intervals (of the order of 1 km\,s$^{-1}$). We kept those very small differences from values in Table \ref{DIB_index} and fixed  $\lambda_{S}$ for the next step. Finally, the coefficient  $\lambda_{D}$ that represents the DIB shift was also found to be the same or nearly the same for a large majority of the stars. This confirms that over such a small field of view the DIB radial velocity variation is very small, both because most of the absorption originates from the first two kpc, and also because the radial motion of the absorbing gas in this direction is small. We then fixed this shift $\lambda_{D}$ at the most frequently found value. For all those four parameters we checked that the outliers were corresponding to low signal-to-noise spectra or to the presence of a spurious feature. We then proceeded to a second fit of all spectra, this time for free $\kappa$ and $\beta$ coefficients only.

%=============

\section{Model adjustments}

\begin{figure}
   \centering
   \includegraphics[width=9cm]{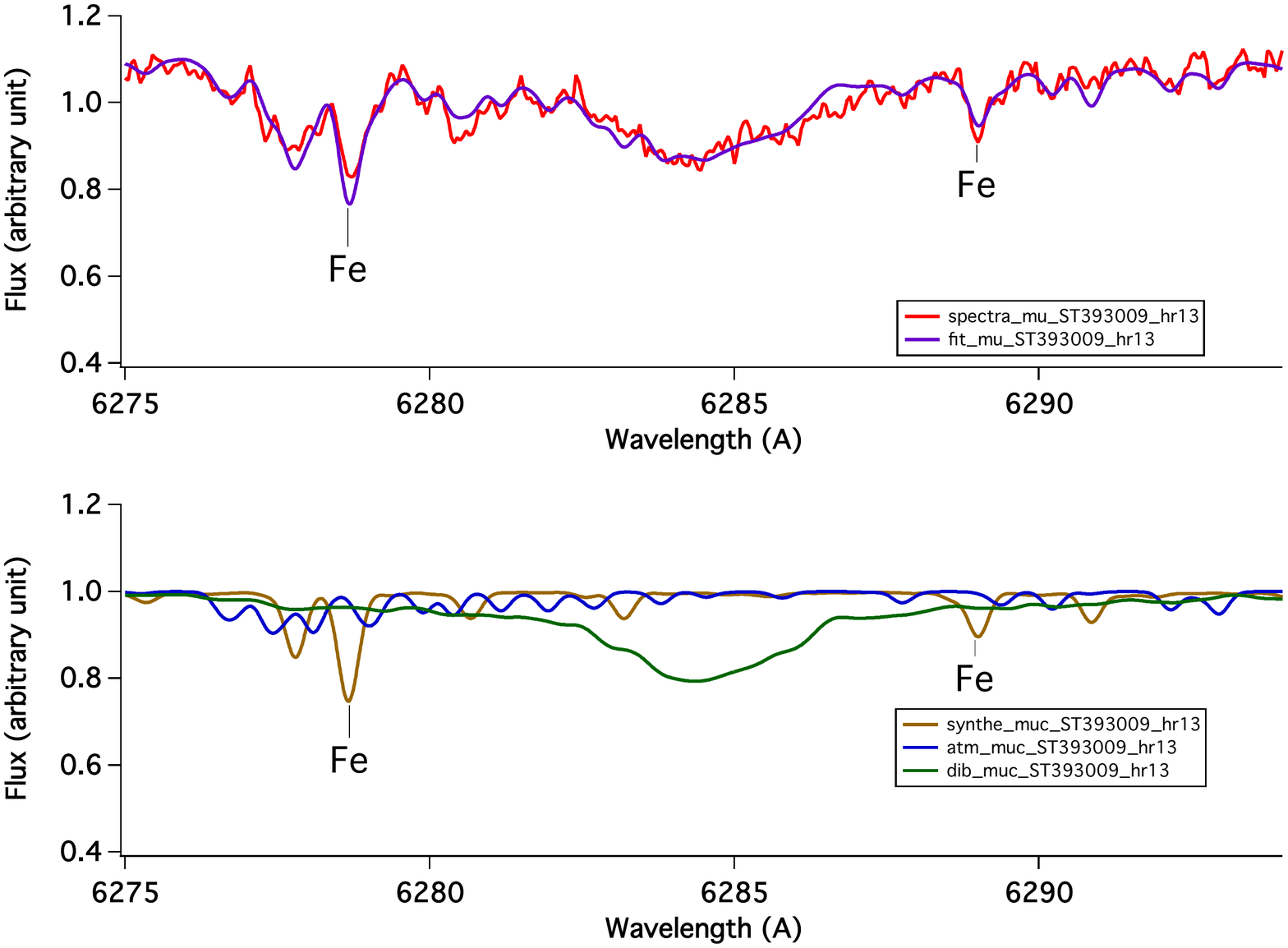}
     \caption{Model adjustment for Ogle N: 393009 (${T_{eff}}$ = 5012 K). The DIB is in a region devoid of strong stellar lines.
     		The upper panel shows the spectrum (red line) and the best-fit model (purple line). 
		The lower panel shows the synthetic stellar model (yellow line), the synthetic atmospheric model (blue line), 
		and the DIB profile (green line) that all correspond to the fit parameters.
              }
         \label{noline}
   \end{figure}

\begin{figure}
   \centering
   \includegraphics[width=9cm]{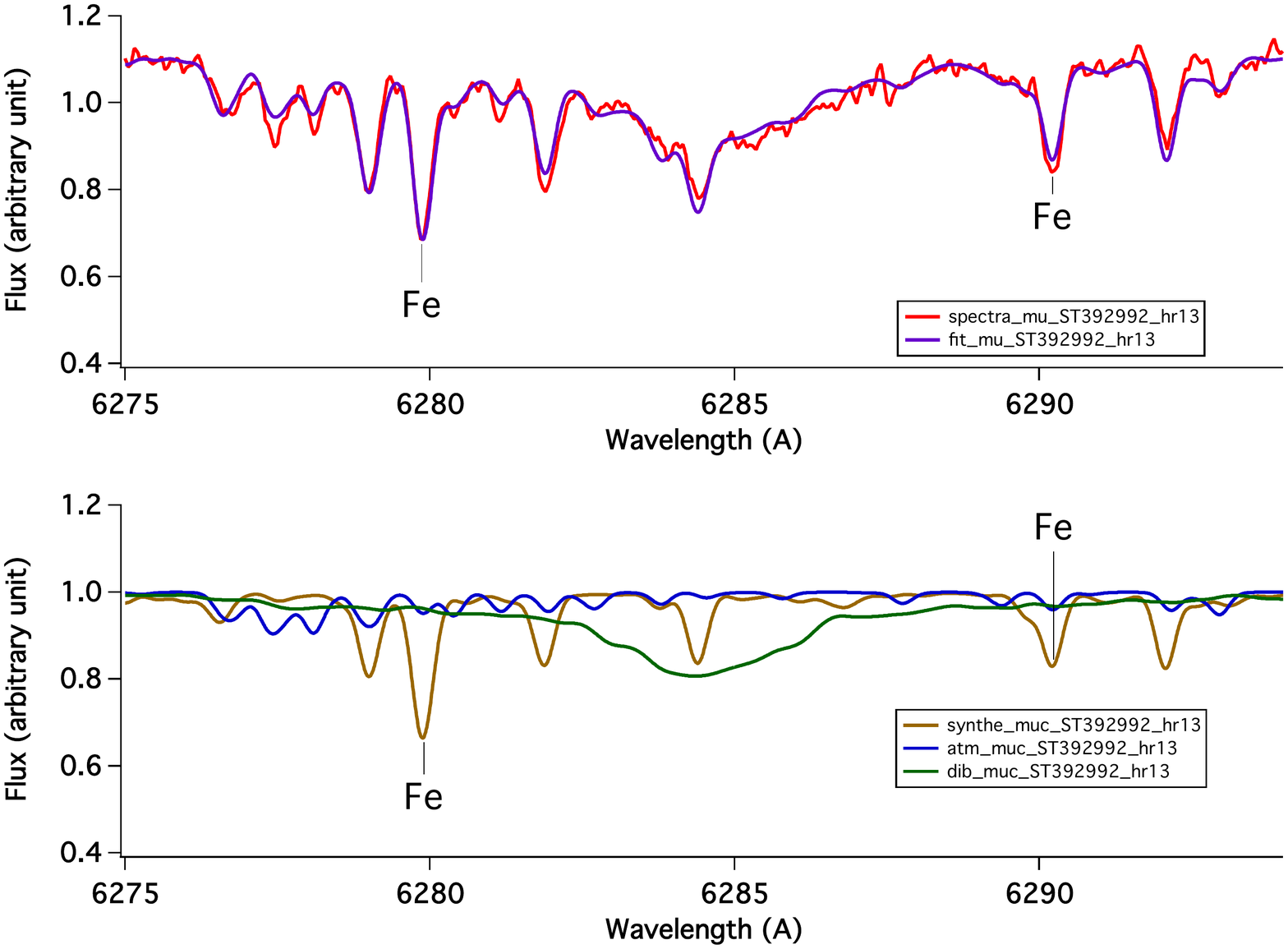}
     \caption{Same as Fig \ref{noline} for  Ogle N: 392992 (${T_{eff}}$ = 4907 K). The DIB here is in a spectral region characterized by moderately strong stellar lines.
        }
         \label{oneline}
   \end{figure}

\begin{figure}
   \centering
   \includegraphics[width=9cm]{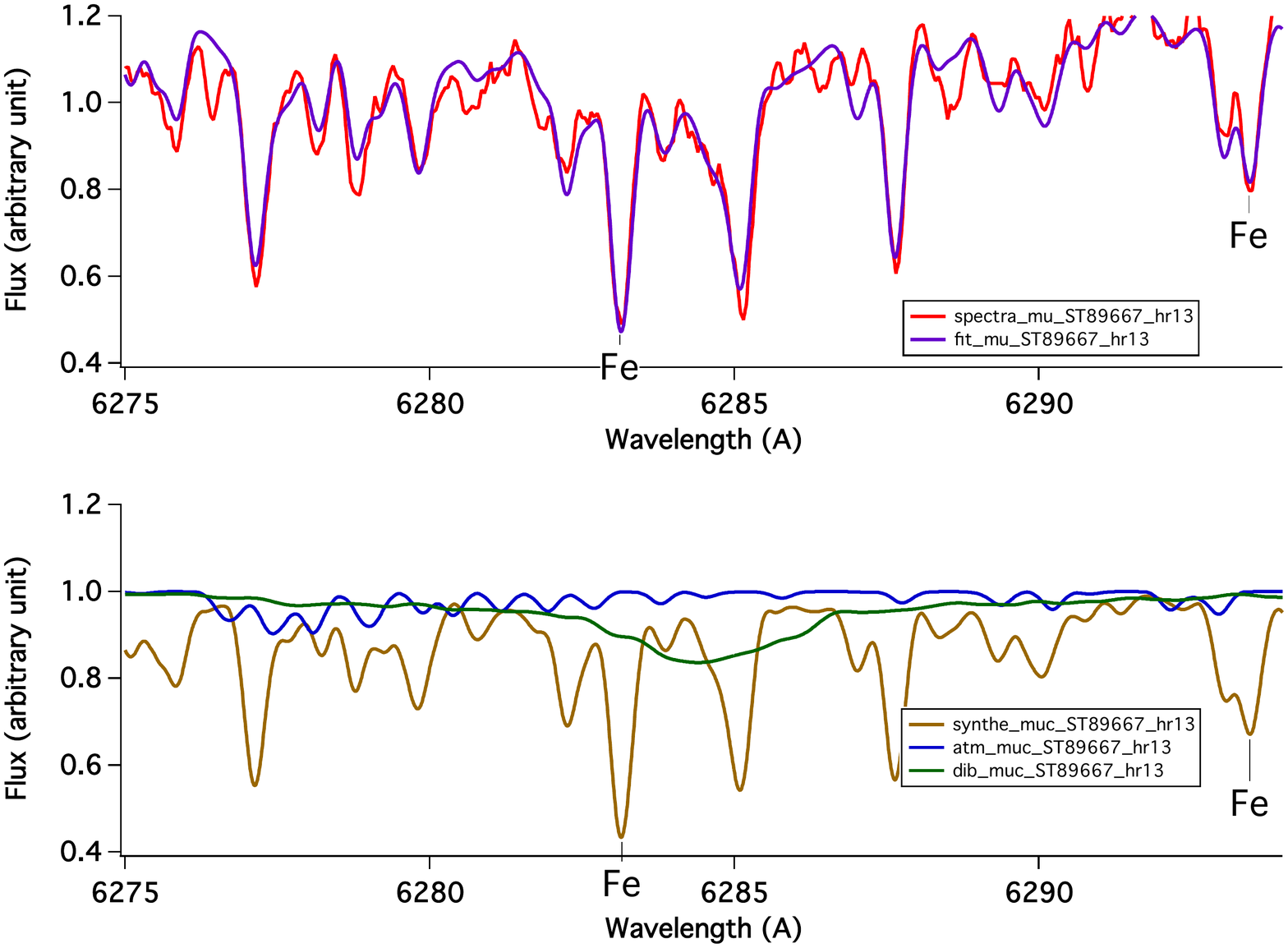}
      \caption{Same as Fig \ref{noline} for Ogle N: 89667 (${T_{eff}}$ = 4516 K). The DIB here is both weak and embedded in  strong stellar lines.
                   }
         \label{weakDIB}
   \end{figure}

Figures \ref{noline}, \ref{oneline} and \ref{weakDIB} illustrate exemplary cases of model adjustments for the three 6196.0, 6204.5, and 6283.8 $\AA$ DIBs chosen for this study. In each figure the lower panel shows the model components after optimization of the parameters, i.e.  the synthetic stellar model (yellow line), the synthetic atmospheric model (blue line), 
and the adjusted DIB profile (green line). The upper panel shows the spectra (red line) and the fitting result (purple line). 
What we aim at showing in those figures are the strong differences linked to the target star individual radial velocities, that vary over a very large velocity interval, from -230 to +250 km.s$^{-1}$. This results in very different locations of the DIBs with respect to the main stellar lines. There are also very large differences between the 6283.8 $\AA$ DIB and the two narrow ones, with a much stronger impact of the star radial velocity and of the signal to noise on the DIB detection for the latter two DIBs. Finally, the larger the metallicity, the stronger the stellar lines and their impact on the DIB measurement, especially again in the case of the two narrow and weak DIBs. 

Figures \ref{noline} to \ref{weakDIB} show three examples of adjustments for the strongest and broadest DIB at 6283.8 $\AA$, 
The first star (Fig. \ref{noline}) corresponds to an optimal case: given the star radial velocity, only weak stellar lines are located in the DIB region, and the fitting program could easily adjust the DIB strength. Moreover, the star is metal-poor (Fe/H = -0.68), which further helps the fitting procedure. Figures \ref{oneline} and \ref{weakDIB} correspond to more difficult cases, with one or more strong stellar lines in the DIB region. In those cases, provided the stellar models adequately predict the stellar line depths, there should not be difficulties in measuring the DIB strengths with accuracy. However, it can be seen that there are for some spectral regions significant and often systematic differences between the observed and modeled lines. Such departures may have several origins:  
the oscillator strengths and other atomic data
(wavelengths, damping constants) that we use have an error attached
to them,  some of the lines are 
simply not  identified and absent from our line list,
our spectra are computed assuming Local Thermodynamic equilibrium (LTE),
departures from LTE in the stellar atmosphere may change the line strength, 
for a given abundance, the model atmospheres we used are one dimensional
static and plane parallel, hydrodynamic effects (granulation)  
may also have effects both on the line strengths and shapes. 
In practice our computed stellar spectrum differs from
the observed spectrum due to a number of shortcomings in 
our modelling.
For example, significant residuals are found at the locations of unidentified lines of the solar spectrum, like the 6273.949, 6282.816, 6286.142, and 6288.315  ${\AA}$ lines (Moore et al. 1966).
Obviously such unidentified lines are not present in our line-list.
Details on these discrepancies can be found in the Appendix, that aims at estimating the uncertainties on the DIBs equivalent widths and empirically correct for systematic effects. Nevertheless, 
despite the observed departures from the stellar model, the DIB is here broad enough for a reliable DIB estimate. Figure \ref{weakDIB} illustrates this property, and shows one of the worst cases of overlapping stellar lines, a metal-rich (although moderately, Fe/H=0.67) target star, and one of the smallest DIB equivalent widths. 
The DIB strength  could be measured reasonably well despite those conditions, 
thanks to the good signal to noise ratio of the spectrum and the width of the DIB.

\begin{figure}
   \centering
   \includegraphics[width=9cm]{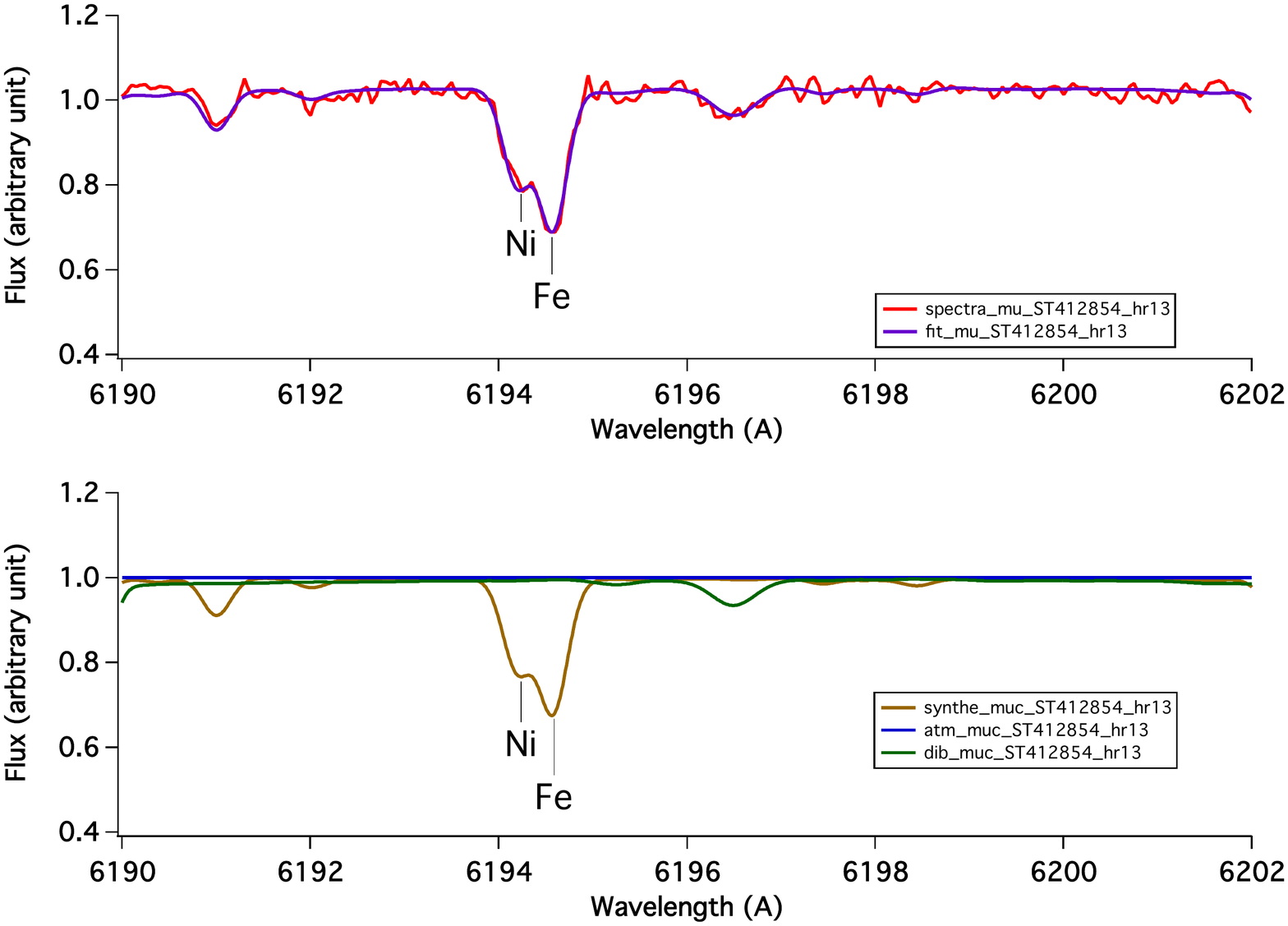}
     \caption{Same as Fig \ref{noline} for the 6196 \AA\ DIB and Ogle N: 412854 (${T_{eff}}$ = 5191 K). The narrow DIB is in a region devoid of strong stellar lines.
              }
         \label{DIB6196A_good}
   \end{figure}

\begin{figure}
   \centering
   \includegraphics[width=9cm]{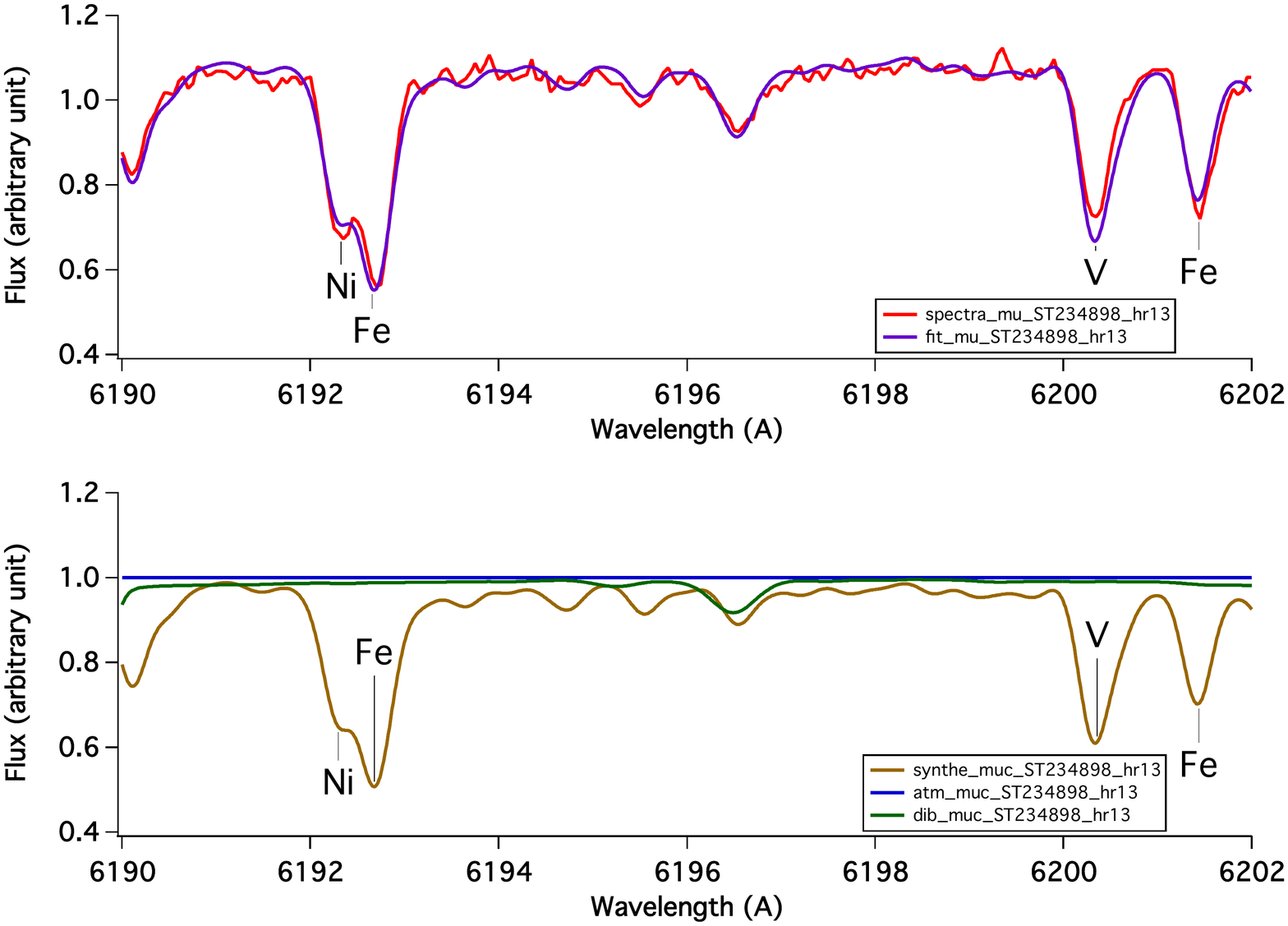}
     \caption{Same a Fig. \ref{DIB6196A_good}  for  Ogle N: 234898 (${T_{eff}}$ = 4714 K). The DIB and a strong stellar line do overlap. }
         \label{DIB6196A_bad}
   \end{figure}

\begin{figure}
   \centering
   \includegraphics[width=9cm]{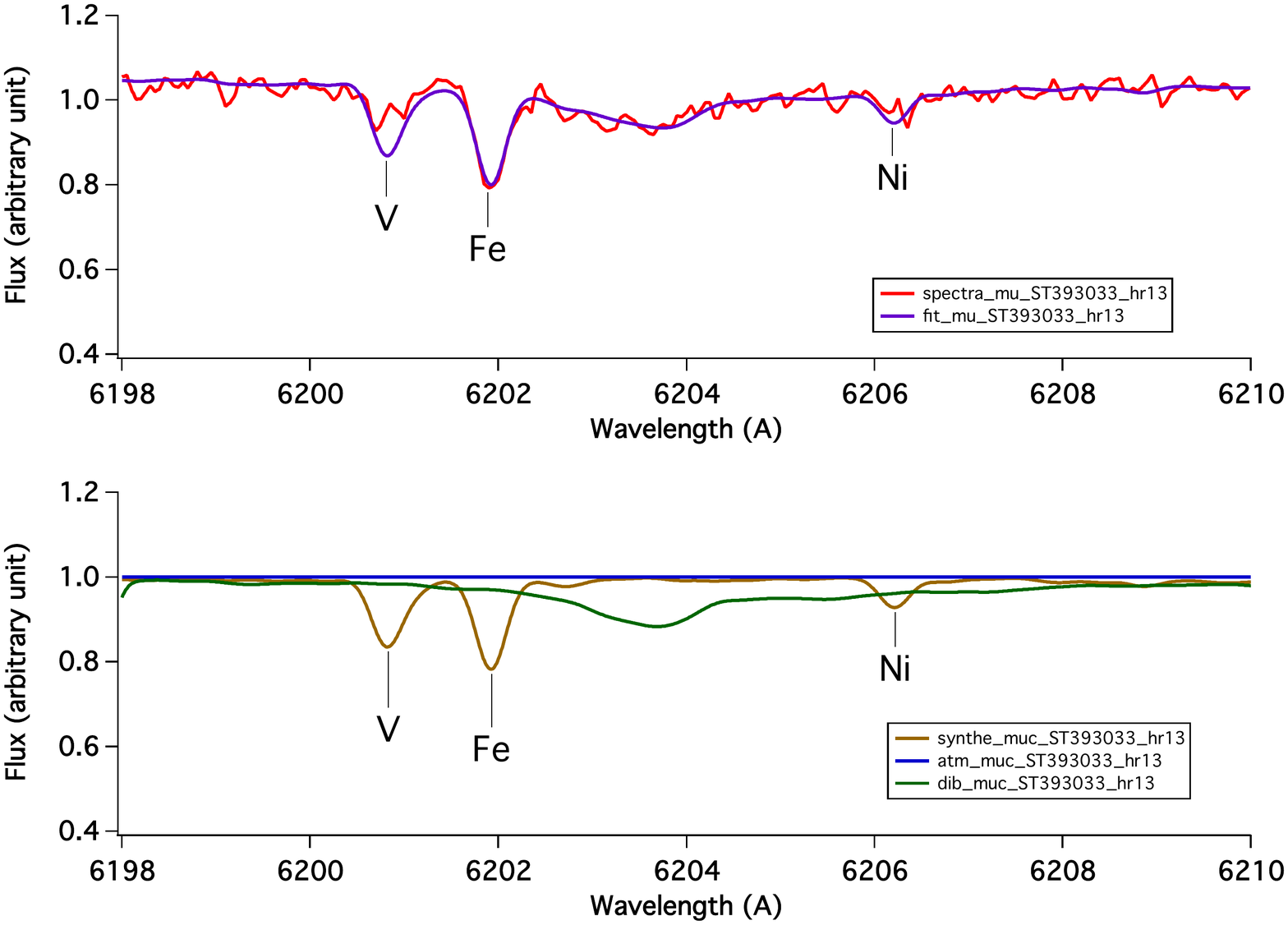}
     \caption{Same as Fig \ref{noline} for the 6204 \AA\  DIB and  Ogle N: 393033 (${T_{eff}}$ = 4914 K). The narrow DIB is in a region devoid of strong stellar lines.
              }
         \label{DIB6204A_good}
   \end{figure}

\begin{figure}
   \centering
   \includegraphics[width=9cm]{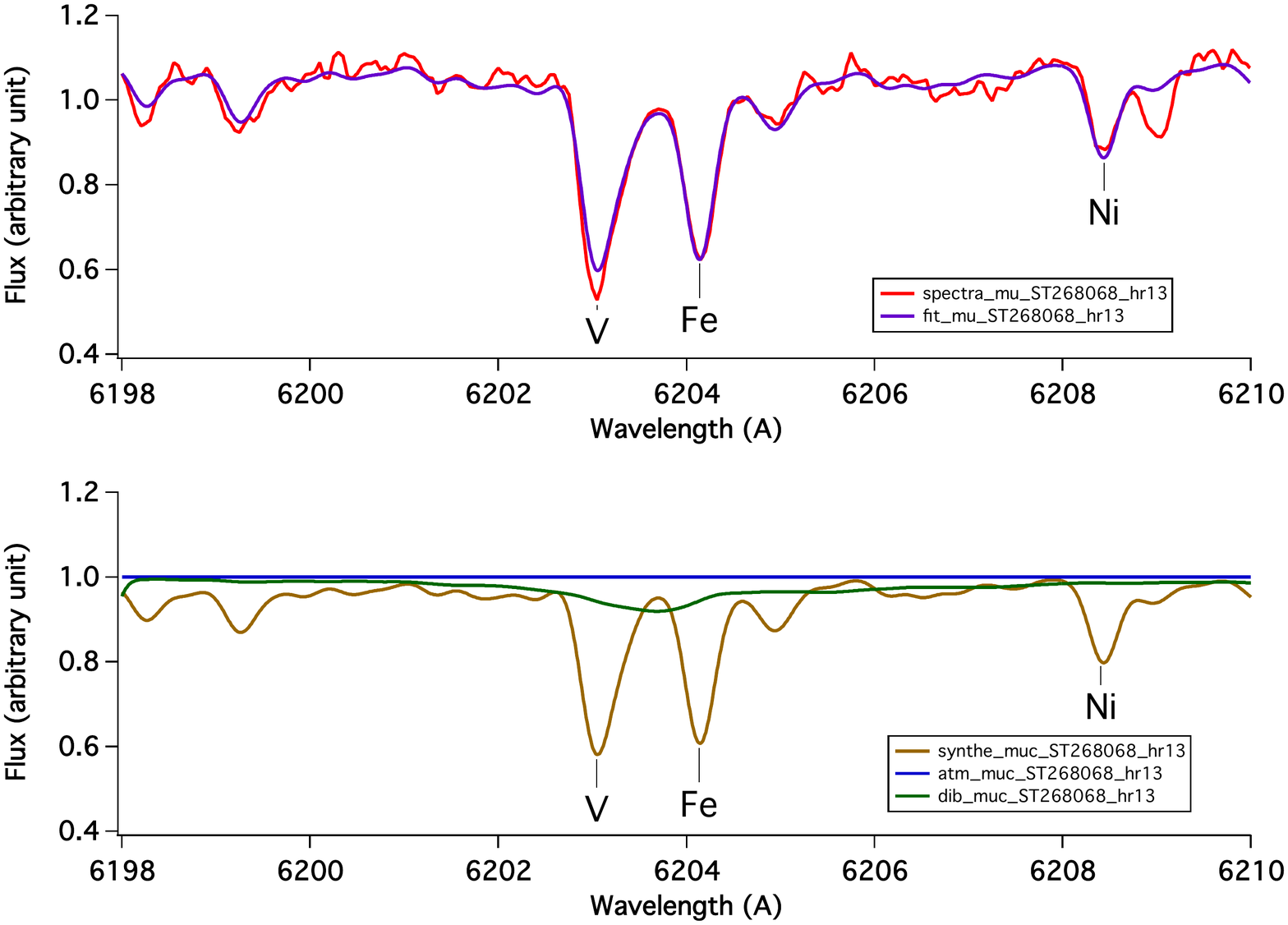}
     \caption{Same as Fig \ref{DIB6204A_good} for  Ogle N: 268068 (${T_{eff}}$ = 4837 K). The DIB region corresponds to strong stellar line.
              }
         \label{DIB6204A_bad}
   \end{figure}

This is not the case for the two narrow DIBs. Figures \ref{DIB6196A_good} to \ref{DIB6204A_bad} show two examples of determinations for each of the  6196.0 and 6204.0 \AA\ DIBs, a relatively {\it easy} one with only weak  stellar lines contaminating the DIB region, and a {\it difficult} one with stellar lines overlapping the DIB. In the first case, it can be seen that DIB equivalent widths can be safely measured. In the second case, it is clear from the figure that the stellar line accuracy is critical, and that some of the measurements  are very uncertain, at least for the low extinction (and DIB strength) regime that prevails here. In comparison with those uncertainties, additional errors linked to the use of a predefined shape of the DIB as well as  to the  telluric model (for the 6284 \AA\ DIB) are negligible. Those systematic departures have been studied in detail and are discussed in the Appendix. Based on them, a first order empirical correction was devised and was applied to the DIB derivation. In brief, residuals for all spectra were all shifted to the stellar frame and the resulting spectra were sorted as a function of the stellar metallicity (see Figure 1 from the Appendix). At each wavelength an average linear relationship between the residual value and the metallicity was adjusted, providing a systematic offset as a function of metallicity and wavelength (Fig 2 and 3 in the Appendix). This offset is maximal at the locations of the over- or under-predicted lines, and null elsewhere. Such an offset was then applied as a corrective term at all wavelengths to all DIB spectra (Fig A-4), and a new adjustment of the model and subsequent computation of the DIB equivalent width were performed after those corrections.  We compared the DIB-DIB relationships and also the DIB-extinction relationships both before and after the empirical correction, and found a systematically better DIB-DIB relationship, with in particular a factor of two increase of the Pearson correlation coefficient in the case of the 6196 \AA\ vs 6204 \AA\ DIB comparison. We also found a significant improvement of the DIB-extinction relationship, except for the 6284 \AA\ DIB for which there was no change. We are conscious that a more fundamental approach would be desirable, but, in view of those improvements, we kept the corrected values for the remaining part of the analysis. More improvements are expected in future from elaborated studies of the stellar spectra in the DIB spectral regions, i.e. individual adjustments of the log(gf), studies of the missing lines, non-LTE and granulation effects.
Such studies are beyond the scope of this work, that is devoted to tests of the new method adapted to cool stars.

\section{DIB equivalent widths, DIB-DIB and DIB-extinction correlations}

Table \ref{DIB_index} lists the resulting equivalent widths for the three DIBs, both before and after the empirical correction, as well as the associated uncertainty. Uncertainties, whose derivations are also described in detail in the Appendix, are a combination of random errors associated to the noise level and of errors linked to the use of the three models that remain after application of the above mentioned empirical correction. These quasi-random uncertainties were derived from the whole set of residual vs metallicity curves that were computed for each wavelength and used for the correction of systematics. They were simply taken as the variance of the residuals around the mean relationship (see Fig \ref{residual_6280A} and \ref{resi_FeH_fitting}). This variances includes both the noise and the departure from the empirical {\it ideal} relationship between the metallicity and the stellar lines. It is clear that those derived uncertainties provide an order of magnitude of the errors which has a sense as a mean for all targets, but that individual errors for each target may be larger or smaller. Future work will address this point, once additional studies of the stellar models will be performed.

\begin{figure}
   \centering
   \includegraphics[width=9cm]{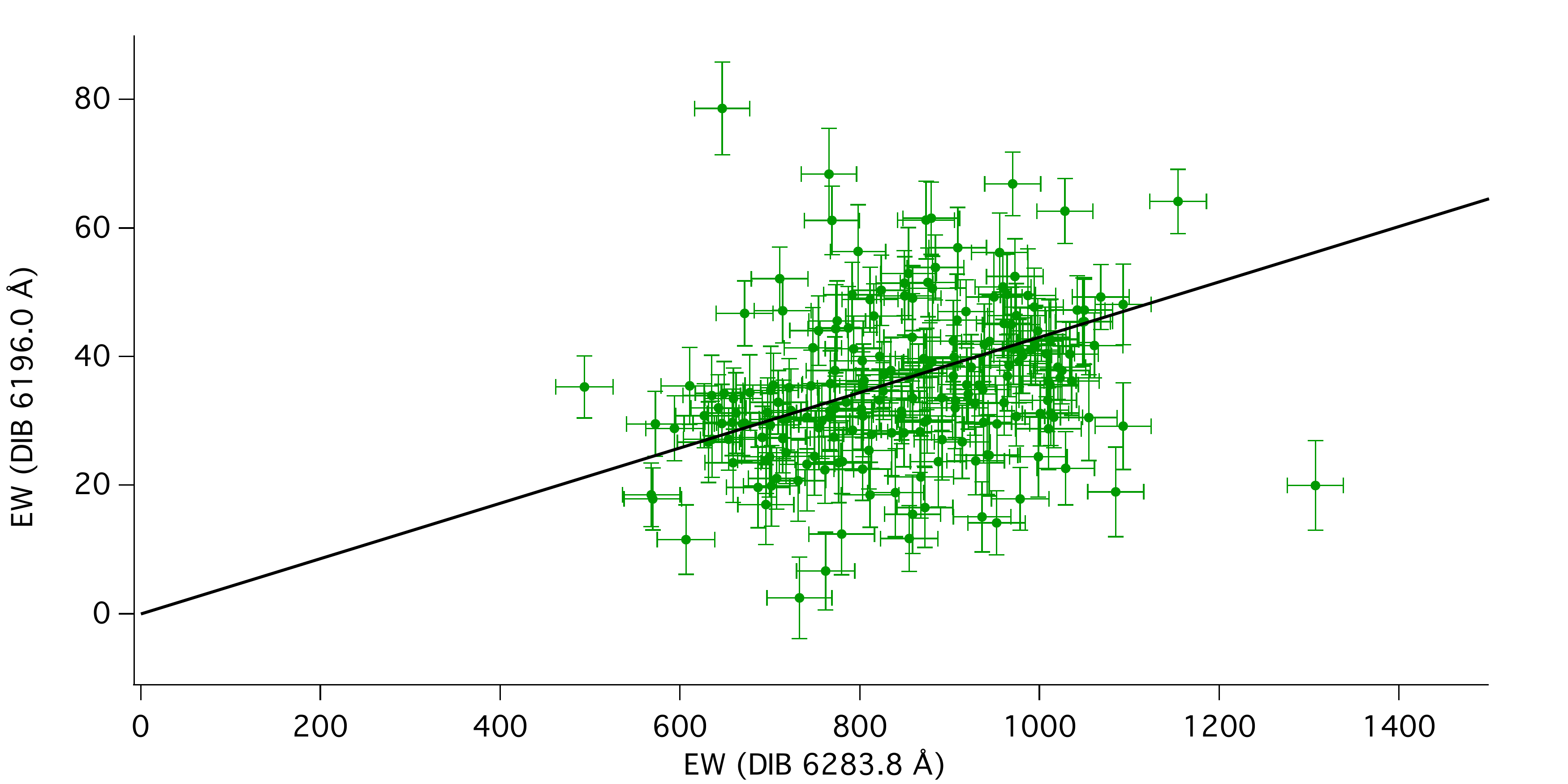}
     \caption{Equivalent width of DIB 6196.0 ${\AA}$ as a function of the equivalent width of  DIB 6283.8 ${\AA}$. The black line is the best linear fit for pure proportionality, using  error bars of both DIBs. The dotted line shows the linear fit with free offset.}
         \label{EW_6196A_6280A}
   \end{figure}

   \begin{figure}
   \centering
   \includegraphics[width=9cm]{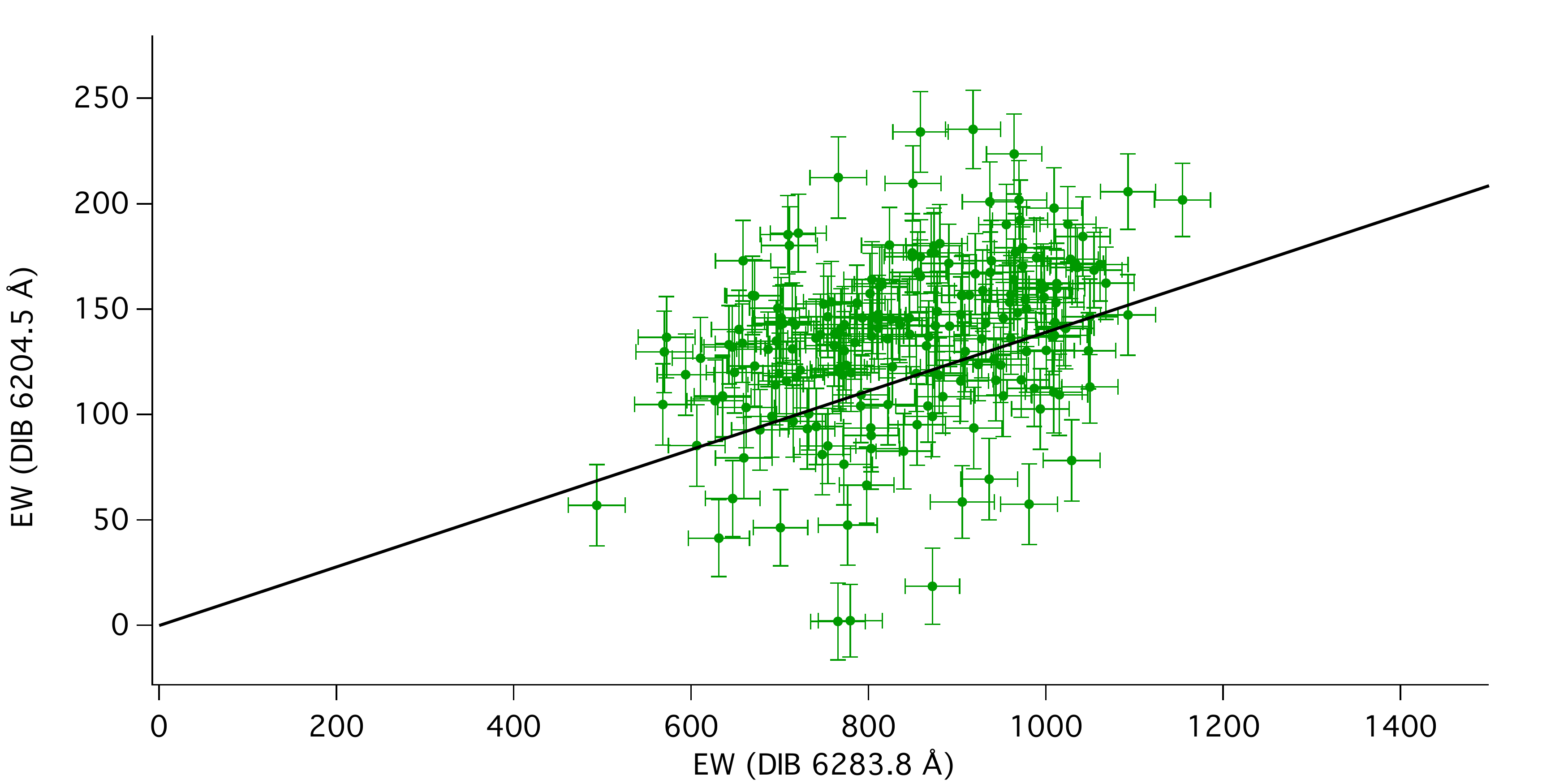}
     \caption{Same as figure \ref{EW_6196A_6280A} for DIB 6204 \AA\ .}
         \label{EW_6204A_6280A}
   \end{figure}
 
\begin{figure}
   \centering
   \includegraphics[width=9cm]{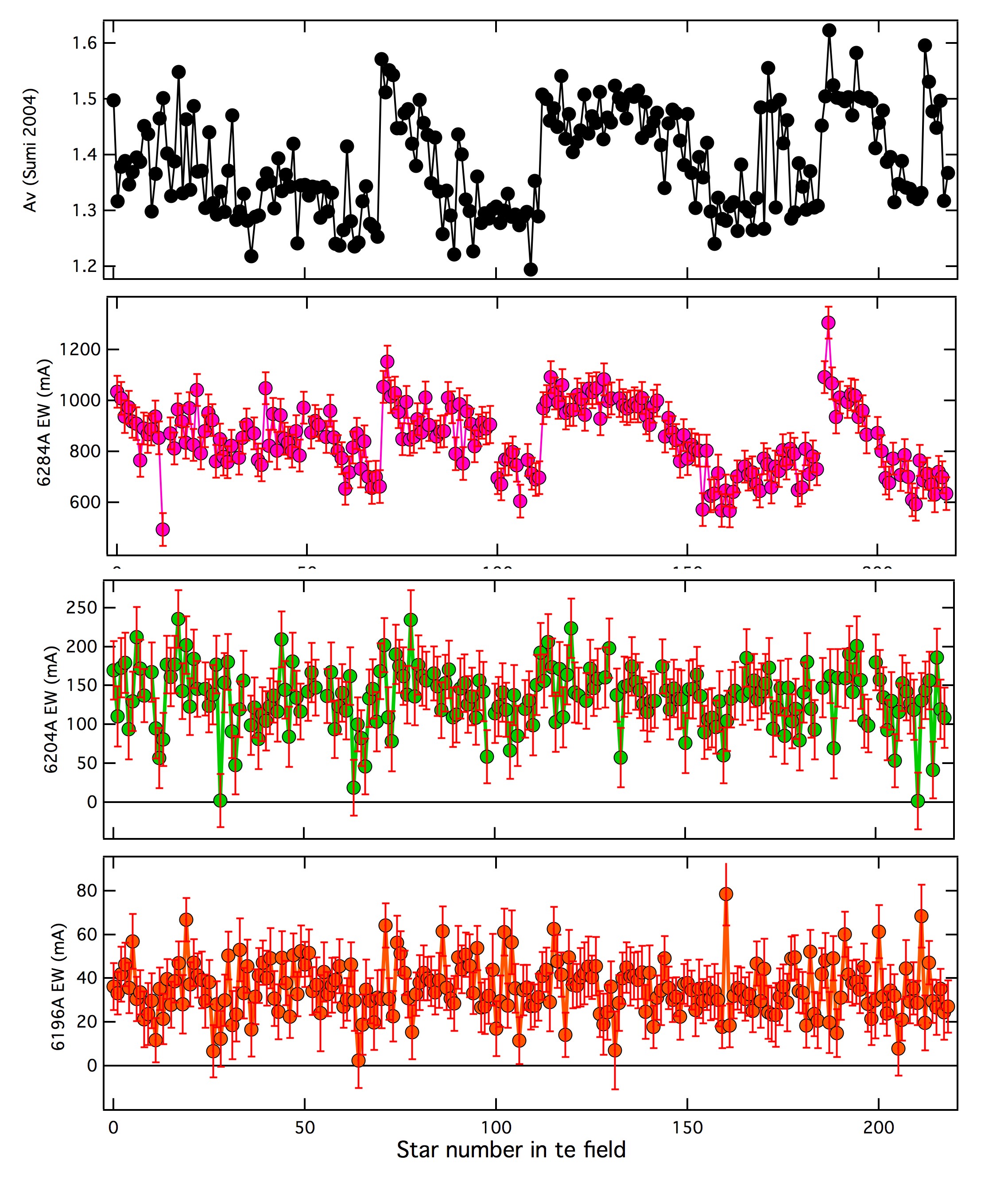}
     \caption{From bottom to top, equivalent widths of the three DIBs (6196, 6204 and 6284 \AA\ ) as a function of the star number
     		and the extinction Av interpolated  from the map of Sumi (2004) at each star location.}
         \label{ew3dibssumi}
   \end{figure}

An immediate test of the reliability of the DIB strengths is the existence of DIB-DIB correlations, because EWs have been measured independently for the three bands. This test is is not so easy here due to the limited range of extinctions (Sumi, 2004). Still, the Spearman's rank correlation statistics allows to reject the absence of two-by-two correlations by better than 99.9 \%.  Figures \ref{EW_6196A_6280A} and \ref{EW_6204A_6280A} show the DIBs 6196.0 and  6204.5  EWs  as a function of the broad 6283.8 $\AA$ EW for the whole sample. The black lines correspond to a pure proportionality. Despite the existence of large uncertainties, especially for the two small DIBs, the correlation between the three bands is clearly visible. 
Those correlations show that the signal we extracted for the DIBs contains some information on the IS absorption, even for the two small DIBs. 
A second test of the DIB  measurement is obtained from a comparison with the extinction map. Fig \ref{ew3dibssumi} displays the star-by-star variations of each  DIB equivalent width as well as the extinction derived from the OGLE photometry obtained by interpolation through the Av map of Sumi (2004). The four patterns, that represent variations across the field of the four independent quantities, reveal large similarities for the strong DIB, some similarities for the 6204 \AA\ DIB. Large uncertainties make the comparison less obvious for the 6196 \AA\ band.  This is reflected in the Spearman's rank correlation statistics that allow to reject the absence of correlation at better than 99.9 \% for the 6284 and 6204 DIBs, and 90 \% for the 6196 \AA\ DIB.

Better statistics could have been obtained after rejections of the noisy spectra or particularly poor adjustments. However, our goal  here is to provide some idea of the DIB extraction results as a function of  the signal quality, the DIB strength and the reddening for an entire field, with an entirely automated method.

\section{Color excess estimates based on nearby star empirical relationships}

DIBs are moderately or only weakly correlated with the extinction, as shown by a number of studies and the existence of DIB {\it families} whose members are internally closely correlated, while distinct families seem to obey different laws (e.g. Krelowski \& Walker 1987, Cami et al. 1997). There is no {\it parental} link between the three DIBs we study here, and thus we do not expect to find strong correlations between their EWs nor very similar dependencies on the extinction.  Previous surveys of stars in the solar neighborhood have provided some statistical relationships between the reddening E(B-V) derived from spectrophotometry and the equivalent width of each DIB. Best-fit parameters of linear correlations have  been established recently by \cite{fri2011} for all three DIBs, based on about 130 O-B northern hemisphere nearby stars within 1 kpc.  In the case of  the 6284 \AA\ DIBs, Raimond et al (2012) derived slightly different coefficients  based on about the same number of southern hemisphere targets, observed with the ESO/La Silla FEROS spectrograph.  Using the same FEROS data, Puspitarini and Lallement (2012) derived linear fit coefficients for the two other DIBs, again found slightly different when compared to the northern survey study. The coefficients for the linear relations between EW and $E_{B-V}$ are shown in Table \ref{EW_DIB_relation} for both surveys and for the three DIBs.
There are some differences between the relationships as they come out from the two surveys, that have been discussed by Raimond et al. (2012). Due to the use of cooler and fainter target stars, the ESO/FEROS DIB strengths are not as much influenced by strong radiation fields and there is significantly less dispersion around the mean DIB-E(B-V) relationship. As a matter of fact, DIB strengths may be significantly reduced in case the main dust cloud responsible for the absorption is very close to the UV-bright target star (Friedman et al. 2011, Vos et al. 2011), an effect attributed to the ionization state change of the carriers in the stellar environment. For this reason, the number of {\it outliers} with a relatively weak DIB and a large reddening is smaller in the FEROS survey, and the dispersion decreases. In parallel, the mean slope E(B-V)/EW(DIB) derived from the sample is also smaller. This was found systematically for the three DIBs. Here we will make use of the FEROS-based mean relationships, because the line-of-sight here is not related to hot bright stars and the DIB should in principle better correspond to average conditions. For comparisons we also show both results for the strong  6284 \AA\ band.
 
 \begin{figure}
   \centering
   \includegraphics[width=9cm,height=8cm]{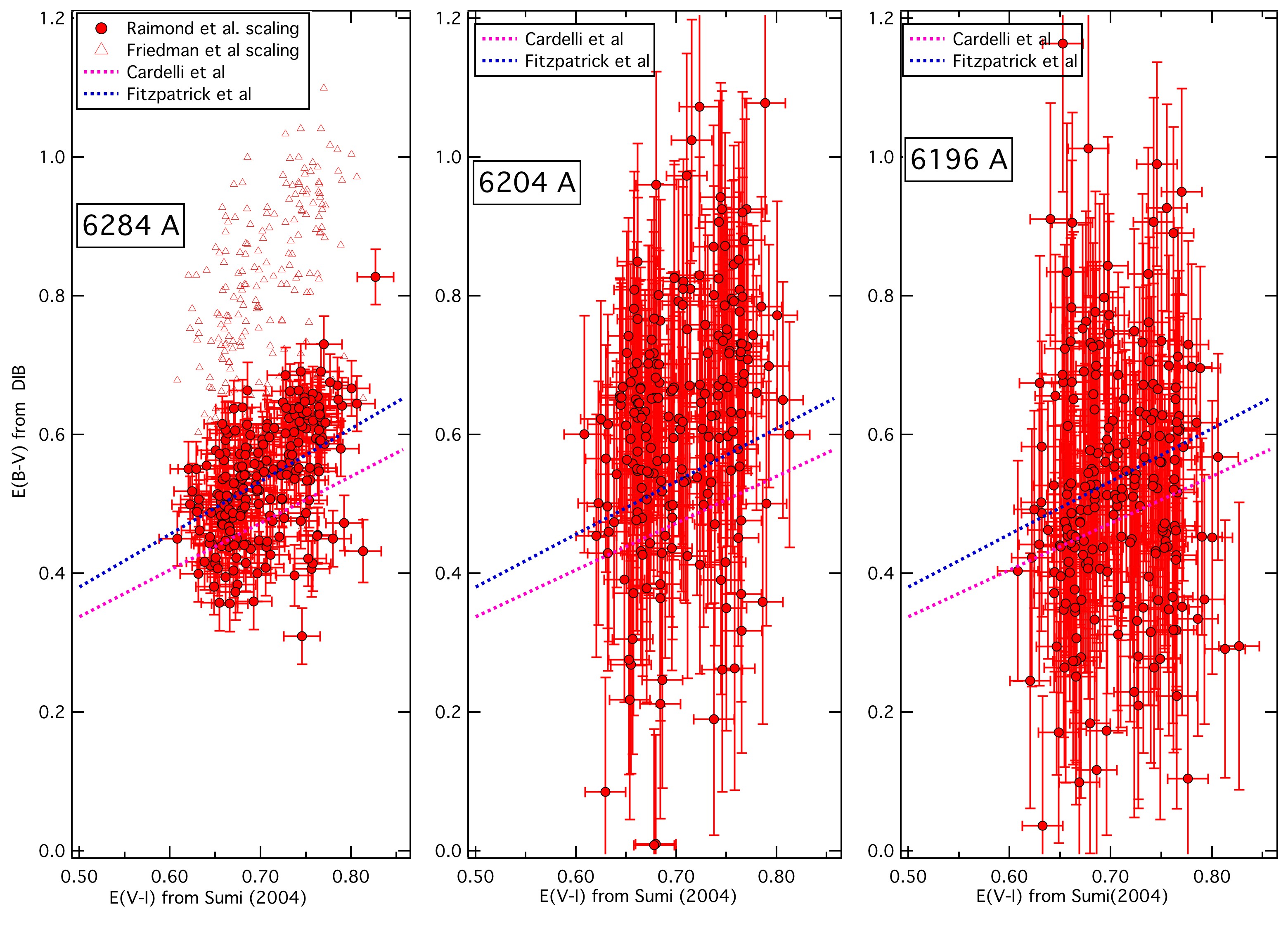}
     \caption{Color excess E(B-V) derived from the three DIBs as a function of the E(V-I) value interpolated within the \cite{sum2004} map at the locations of the target stars. 
     		E(B-V)s here are obtained using the Raimond et al. (2012) and Puspitarini et al. (2012) average relationships based on FEROS data.  The blue and pink dotted lines correspond to color excess relationships based on the Fitzpatrick (1999) and Cardelli et al. (1989) extinction curves and Rv=3.1 (see text). In the case of the 
    DIBs 6283.8 ${\AA}$ we also show the color excess values deduced from  the Friedman et al (2011) average relationships  (open triangles).}
    
 \label{EBVsumi_2rvs.jpg}
\end{figure}

 The measured DIB equivalent widths derived from the global fitting were converted into {$E_{B-V}$} color excess values by application of the three average relationships listed in Table \ref{EW_DIB_relation}. 

Figure \ref{EBVsumi_2rvs.jpg} shows the color excess $E_{B-V}$ estimate based on the  DIB 6283.8 ${\AA}$ and the FEROS relationship, as a function of the color excess $E_{V-I}$ interpolated 
from the \cite{sum2004} maps (x-axis). The figure is somewhat redundant with the previous one, however it provides a  better idea of the equivalent width dispersion and the global relationship. The data point distribution shows a correlation between the DIB-based and photometric determinations, but there is a large dispersion, and also the observed interval for the DIB strength is significantly larger than the range of variation of the photometric determination (the best fit linear relationship does not go through zero). The relationship between $E_{B-V}$  and $E_{V-I}$ depends on the extinction curve, and in less extent on the stellar spectrum. We have drawn here the relationships based on both classical Cardelli et al. (1989) and Fitzpatrick (1999) extinction curves, for Rv= 3.1.  E(V-I) is computed for a typical red-clump giant star, .and E(B-V) is computed for the A0 star Vega. As far as average absolute values are concerned, the DIB-based $E_{B-V}$s are similar to what one would expect from those classical laws, and in better agreement with the Fitzpatrick relationship.   We also show the $E_{B-V}$ values we would obtain with the empirical Friedman et al (2011) relationship. Those $E_{B-V}$ empirical estimates are systematically higher and more difficult to reconcile with the photometric data. We believe that the relationship based on the cooler FEROS targets maybe more appropriate here because it probes environmental conditions that are closer to the average conditions encountered here than to the high ionization conditions encountered towards O- and early B-type nearby target stars.

Figure  \ref{EBVsumi_2rvs.jpg} also shows how the 6196 A and 6204 A DIB-based color excess $E_{B-V}$ compares with the photometric determination of $E_{V-I}$. For those two DIBs, the correlation is visible, and the agreement between the measured and the expected ratio between the two color excess values is again better with the Fitzpatrick et al. reddening law. There appears to be less deviation from average simple proportionality for those DIBs, compared to 6284\AA\ , especially for the DIB 6196 for which the best-fit relationship is compatible with proportionality within the uncertainty range.

 An important source of dispersion is the fact that the Sumi color excess is an average over the series of stars contained within the 0.6 x 0.6 arcmin pixel field and located at various distances in the bulge, while the DIB is derived for individual stars.  Note that here the main effect would be angular variability, because, as said above, the quasi-totality of the extinction is generated closer than 2,000 pc, thus differences in the distances to the bulge targets have in principle no impact.  At 1 kpc, 0.6 arcmin corresponds to a transverse distance of 0.17 pc, thus only very dense, small cloud cores could produce variations as large as those observed.
 A likely explanation for both the dispersion and also departures from proportionality between the DIB-based color excess and the photometric E(V-I) is the existence of intrinsic DIB variations in response to the radiation field and in general to environmental conditions, such like those found at smaller distances. This is favored by the spatial pattern as we will discussed below. On the other hand, spatial variations of the DIB-E(V-I) ratio may also be related to spatial variability of the extinction curve. Udalski (2003) and Sumi (2004) both  consider that towards the bulge Rv may vary significantly. A relationship between the extinction curve, the total to selective extinction ratio and the DIB behavior has been first discussed by Krelowski et al (1999). Environmental effects and dust processing on one hand, and extinction curve    
on the other hand are themselves related through the grain distribution and the cloud history, which also influences DIB carriers and strengths.

\begin{figure}
   \centering
   \includegraphics[width=9cm]{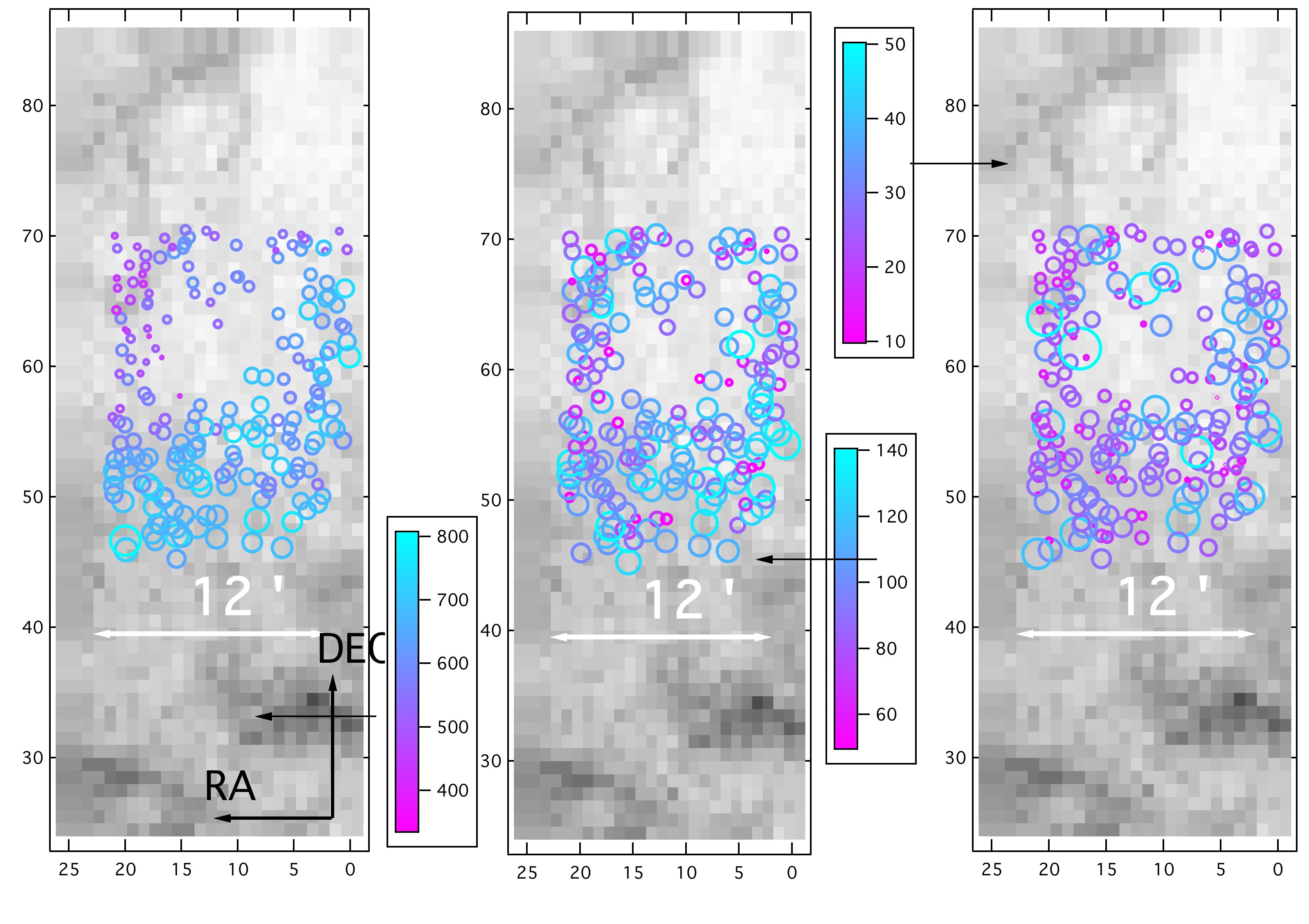}
     \caption{Spatial variability of the DIB and the DIB-extinction ratio (from left to right, the 6284,  6204, 6196 \AA\ DIBs). Target stars used in this study are superimposed on the Av extinction map from Sumi (2004) (grey scale). The size of the circle is proportional to the DIB equivalent width, and its color corresponds to the ratio  between the  DIB equivalent width and the extinction Av obtained by interpolation through the Sumi map (color scale).}
         \label{Layout2gray}
   \end{figure}

The spatial variability of the ratio between DIB-based and photometric determinations of the color excess is shown superimposed on the Sumi map in Figure \ref{Layout2gray}. In this figure the color indicates the ratio and the size of the markers is proportional to the  DIB strength. The distribution of colors shows that there are some systematic changes across the field for the 6284 DIB, in particular the DIB is relatively small compared to the photometric value at larger declinations and right ascensions. Future investigations will hopefully allow to attribute or not those differences to variations of the environmental conditions in the encountered clouds, or to extinction law spatial variations, or both, as discussed above.  The 6204 \AA\ DIB shows the same behavior than the 6284 band, although less clearly due to large uncertainties and also not so pronounced. At variance with the two other bands, the 6196 \AA\ DIB does not reveal any spatial trend, although there are strong uncertainties on the DIB equivalent width.
If confirmed, for the three DIBS the spatial homogeneity of the measured ratio follows the degree of correlation of the DIBs with the color excess, as measured in the solar neighborhood and for widely distributed targets by Friedman et al (2011). The correlation degree and spatial homogeneity decrease from the 6284 \AA\ DIB to the small 6196 \AA\ DIB . 

\begin{figure}
   \centering
   \includegraphics[width=8cm]{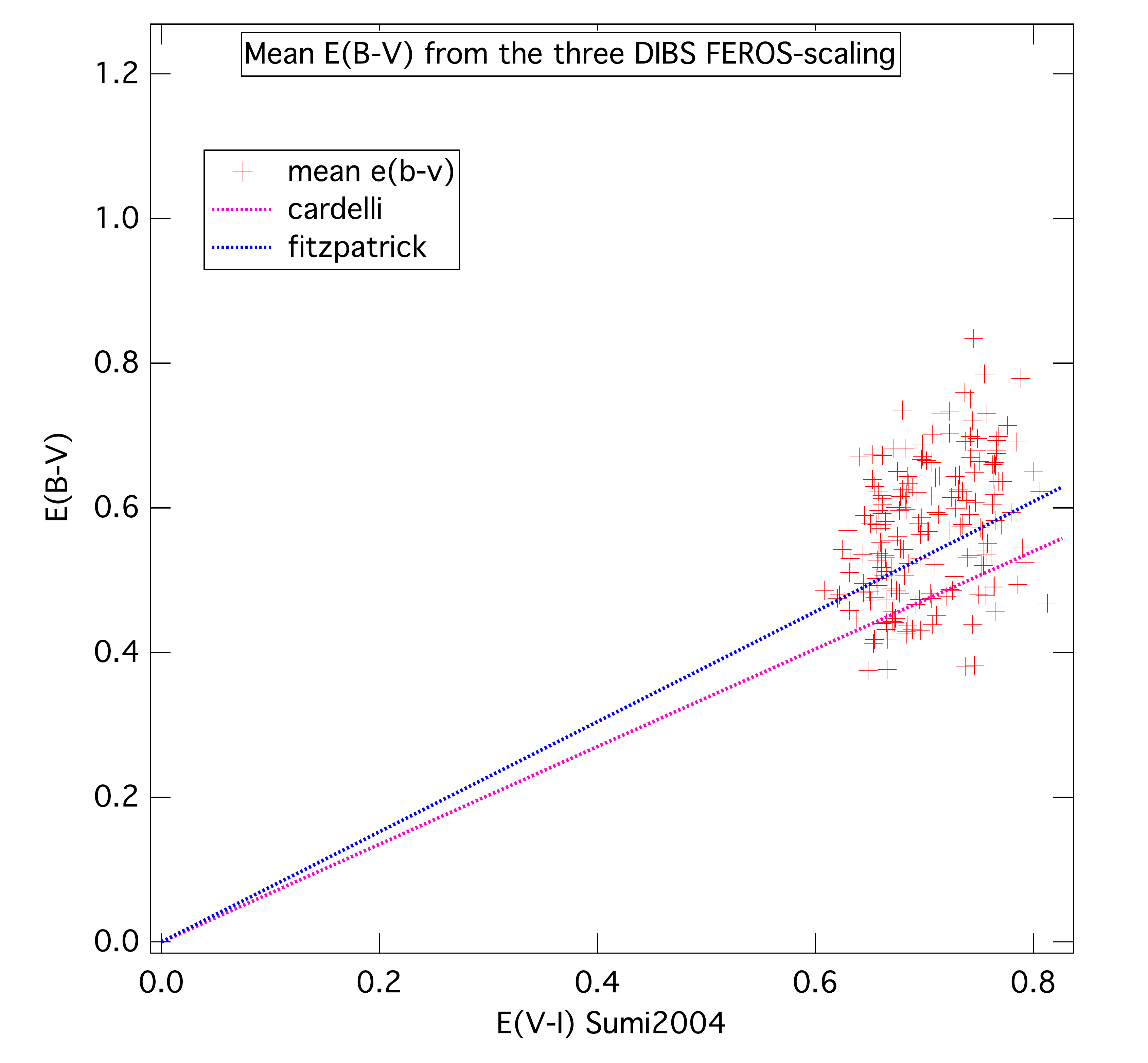}
     \caption{A synthesis of all E(B-V)s estimates from the 3 DIBs, under the form of a weighted mean, here compared with the E(V-I) determination from the OGLE photometry. The blue and pink dotted lines correspond to color excess relationships based on the Fitzpatrick et al. and Cardelli et al. extinction curves.}
         \label{allebvs}
   \end{figure}

Finally, we show in Figure \ref{allebvs} a weighted mean value of $E_{B-V}$ derived from the three DIBs, again compared with the $E_{V-I}$ value deduced from the OGLE analysis. This figure allows to figure out to which extent the combination of those three DIBS can be used here as a first estimator for the reddening, in the absence of photometric measurements. The average linear relationship is $E_{B-V}$= (0.72 $\pm$ 0.14) $E_{V-I}$ + (0.06 $\pm$0.10) and the standard deviation is 0.08, i.e. of the order of 15\% of the average value. This should be compared with other conditions and distances.

\begin{table}
\caption{ E(B-V)$_{\lambda}$=a+b*EW$_{\lambda}$.
	Reference 1: Friedman et al. \cite{fri2011}; 2: Puspitarini and Lallement (2012); 3: Raimond et al. \cite{rai2012}.}
\label{EW_DIB_relation} 
\centering                          
\begin{tabular}{c c c c}  
%\hline
DIB$_{\lambda}$ & Reference & a & b\\ 
\hline                        
   6196.0 & 1 & (-5.07$\pm$0.56)*$10^{-2}$ & (2.11$\pm$0.02)*$10^{-2}$ \\
   	      &  2 & 0					    & (1.48$\pm$0.09)*$10^{-2}$ \\
   6204.5 & 1 & (-7.22$\pm$0.67)*$10^{-2}$ & (5.99$\pm$0.08)*$10^{-3}$ \\
   	      &  2 & 0					    & (4.58$\pm$0.37)*$10^{-3}$ \\
   6283.8 & 1 & (-7.71$\pm$0.78)*$10^{-2}$ & (9.57$\pm$0.17)*$10^{-4}$ \\
   	      & 3 & (-0.50$\pm$0.40)*$10^{-2}$ & (6.37$\pm$0.23)*$10^{-4}$ \\
\hline  
\end{tabular}
\end{table}

\section{Summary and discussion}

We have used R=22,500, S/N= 30-77 observations of 219 red clump giants from the galactic bulge in the Baade's Window (Av $\simeq$ 1.4) as a test case of our newly developed composite model and automated fitting method of interstellar absorption extraction  from cool star spectra. 
The combination of synthetic stellar models, synthetic atmospheric lines, and DIB profiles allowed to extract equivalent width values for the three  6284, 6196 and 6204 \AA\  DIBs. The existence of DIB-DIB correlations demonstrates that, even without any adaptation of the stellar model, in the case of a moderate color excess  E(B-V) $\simeq$ 0.4 and signal to noise ratios above 30, a strong DIB like the 6284 \AA\  band but also narrow and weaker DIBs like the 6196 or the 6204 \AA\ bands can be measured for cool stars, provided one takes into account the velocity shift between the star and the absorbing ISM. 
More precisely, all spectra could be efficiently used in the case of the strong and broad 6284 $\AA$ DIB (width $\simeq$ 3.5 \AA\, depth $\simeq$ 20\%), while for the two narrow and weaker DIBs ($\simeq$ 0.5 and 1 \AA\ width, 10 \% depth) the potential extraction of the DIB equivalent width depends on the star's radial velocity, whose value results or not in an overlap of stellar lines and DIBs. This is due to the presence of unidentified lines as well as over- or under-predicted line strengths. Model improvements are beyond the scope of this work, but should be performed in future, once other analyses confirm the present trends and enough constraints are obtained. Here we simply performed an empirical correction, independently for each DIB, that is described in the Appendix. Its validity was demonstrated by a strong improvement of the DIB-DIB correlations as well as  improvements of the DIB-extinction correlations. More work is needed on a more fundamental approach to those corrections. Globally, this modeling demonstrates that DIBs can be measured in an automated way for large number of cool targets during spectroscopic surveys, and be used as any other IS line to locate the IS matter.

For the three DIBs the spatial pattern globally reflects extinction variations deduced from stellar photometry, which also validates the fitting method. The degree of correlation is, as expected, globally better for the broader and stronger DIB because uncertainties on the equivalent widths are much smaller. DIB strengths are converted into color excesses, using best-fit linear relationships established for early-type star surveys. Mean values of E(B-V) over all targets are 0.53, 0.62 and 0.52  for the 6284, 6204, and 6196 \AA\ DIBs respectively.  Discrepancies among the DIBS amount to about 20\%, and there is a 15\% dispersion of the weighted mean value around its average linear relationship with the photometric reddening (Fig. \ref{allebvs}). 
Several results were obtained from the analysis, that call for further studies. The 6284, and in a smaller extent the 6204  \AA\ DIBs  amplitude intervals over the field are found to be larger than the color excess relative variations deduced from OGLE photometric data and their analysis (Sumi 2004, see Fig \ref{ew3dibssumi}).  This is reflected in a spatial variability of the DIB-based to photometry-based color excess ratio.  Those variations are not randomly located, as is especially visible for the 6284 \AA\ DIB: the DIB is systematically weaker around $\alpha,\delta$ = 270.95,-30 $^{\circ}$. 
 One likely explanation for such departures from proportionality is the DIB response to ionization conditions in the clouds, and other environmental effects such like shocks.This may be linked to a spatial variability of the extinction law. The grain size distribution influences the shape of the extinction law and he total to selective extinction ratio, but is also linked to the cloud physical properties and its history, that themselves influence the quantity of macro-molecules and the DIBs. Such a link between the total to selective extinction ratio and the DIBs has been discussed by Krelowski et al. (1999).
The excess of amplitude variation and the spatial variability do not seem to exist for the small 6196 \AA\  DIB, although large uncertainties make the comparison with the photometric extinction more difficult. This may be put in relation with the fact that this DIB has been found to  be better correlated with the extinction than the two others, something again potentially related to its different response to the dust distribution, or the environmental conditions. Moreover, the absolute value of the extinction based on the 6196 \AA\ DIB  (again deduced from the average solar neighborhood relationship) seems to agree better with the photometric determination as compared with the two other DIBs.

More analyses should be performed over various fields and for targets at various distances to confirm unidentified or poorly predicted stellar lines. More data should help refining the stellar models in the DIB spectral regions and subsequently improving the DIB extraction. On the other hand,  extinction estimates would certainly be strongly improved by the use of multiple DIBs and of their ratios, accompanied by a better understanding of their specific behavior. Such studies may also provide additional information on the links between the extinction law and the DIBs.

\begin{acknowledgements}
We thank the referee for his useful comments on the manuscript. H-C C. wants to acknowledge the Taiwanese government for her scholarship NSC100-2917-I-564-057. 
\end{acknowledgements}

%=============
\longtab{3}{
\begin{longtable}{crcccrrrrrrrr}
\caption{\label{DIB_index} Stellar data and measurement of EW}\\
\hline\hline
Star & OgleN &  T$_{eff}$ &  log $g$ &  $\xi$ & [Fe/H] &  $E_{B-V}*$ &  EW(6196)&   corr'd &  EW6204)  & corr'd  &  EW6284 &  corr'd  \\
	&    & K	     & dec        & kms$^{-1}$ & dec &     &  m${\AA}$ & m${\AA}$ & m${\AA}$ & m${\AA}$ & m${\AA}$ &m${\AA}$\\
\hline
 \endfirsthead
 \caption[]{(continued)}\\
 \hline\hline
Star & OgleN &  T$_{eff}$ &  log $g$ &  $\xi$ & [Fe/H] &  $E_{B-V}*$ &  EW(6196)&   corr'd &  EW6204)  & corr'd &  EW6284 &  corr'd  \\
	&    & K	     & dec        & kms$^{-1}$ & dec &     &  m${\AA}$ & m${\AA}$ & m${\AA}$ & m${\AA}$ & m${\AA}$ &m${\AA}$\\
 \hline

 \endhead
 \hline

 %\hline
 \endfoot
\footnotetext[1]{* $E_{B-V}$ is derived from the Sumi \cite{sum2004} by interpolation.}
%==============

1&67536&5121&2.43&1.1&-0.45&0.483&40&36$\pm$11&164&170$\pm$38&1046&1036$\pm$62\\
2&67610&4865&2.34&1.5&-0.31&0.425&36&33$\pm$10&110&111$\pm$38&1053&1009$\pm$64\\
3&67641&4575&2.17&1.4&-0.36&0.445&50&42$\pm$13&164&173$\pm$38&962&939$\pm$64\\
4&67663&4878&2.33&1.4&-0.05&0.448&50&46$\pm$10&175&179$\pm$39&1023&974$\pm$64\\
5&67675&4507&2.16&1.5&0.42&0.434&49&36$\pm$12&87&94$\pm$38&991&919$\pm$64\\
6&67687&4664&2.27&1.5&0.54&0.442&64&57$\pm$13&130&130$\pm$38&942&909$\pm$64\\
7&67727&4994&2.45&1.5&0.55&0.45&37&31$\pm$10&206&213$\pm$38&838&766$\pm$64\\
8&67744&5170&2.42&1.4&-0.43&0.447&36&34$\pm$10&175&172$\pm$37&895&891$\pm$62\\
9&67762&4943&2.34&1.3&-0.26&0.468&21&21$\pm$13&131&137$\pm$34&884&868$\pm$73\\
10&78185&4763&2.3&1.2&0.21&0.464&13&24$\pm$14&105& &840&887$\pm$62\\
11&78195&4807&2.34&1.1&-0.05&0.419&34&30$\pm$13&161&167$\pm$38&951&938$\pm$62\\
12&78202&4460&2.17&1.5&0.54&0.441&19&12$\pm$10&88&95$\pm$38&924&855$\pm$64\\
13&78205&5271&2.37&1&0.2&0.473&39&35$\pm$10&53&57$\pm$38&555&493$\pm$64\\
14&78215&4738&2.22&1.6&0.55&0.484&38&22$\pm$12&53&81$\pm$34&747& \\
15&78255&4810&2.31&1.4&0.49&0.452&38&40$\pm$9&184&177$\pm$37&852&871$\pm$63\\
16&78271&4685&2.21&1.5&0.17&0.428&24&28$\pm$11&155&161$\pm$38&816&813$\pm$62\\
17&78281&4899&2.31&1.1&0.2&0.448&32&39$\pm$10&179&177$\pm$37&945&966$\pm$63\\
18&78283&4512&2.21&1.5&0.35&0.5&48&47$\pm$10&242&235$\pm$37&895&918$\pm$63\\
19&78288&5136&2.3&1.5&-0.48&0.429&27&28$\pm$13&126&143$\pm$36&800&835$\pm$63\\
20&78322&4978&2.37&0.8&-0.67&0.472&70&67$\pm$10&205&202$\pm$37&978&970$\pm$63\\
21&78323&4867&2.35&1.5&0.3&0.431&27&37$\pm$10&131&123$\pm$36&799&827$\pm$63\\
22&78330&4595&2.18&1.5&0.61&0.48&40&47$\pm$11&180&185$\pm$37&1035&1042$\pm$62\\
23&78354&4738&2.29&1&0.03&0.442&35&41$\pm$10&152&146$\pm$36&768&793$\pm$63\\
24&78357&4620&2.09&1.5&0.37&0.442&29&39$\pm$13&136& &840&880$\pm$63\\
25&78379&4775&2.31&1.3&0.03&0.421&34&30$\pm$12&131&146$\pm$34&960&952$\pm$73\\
26&78393&4943&2.26&1.2&-0.18&0.465&33&38$\pm$10&133&124$\pm$34&903&924$\pm$63\\
27&78401&4851&2.24&1.5&0.14&0.424&11&7$\pm$12&140&138$\pm$38&818&762$\pm$64\\
28&78421&4632&2.22&1.5&0.46&0.417&25&28$\pm$11&177&177$\pm$37&838&849$\pm$62\\
29&78436&4541&2.19&1.3&0.32&0.43&23&12$\pm$13&-21&2$\pm$34&802&780$\pm$73\\
30&78449&4947&2.22&1.7&0.54&0.418&25&30$\pm$12&155&154$\pm$38&759&758$\pm$62\\
31&78461&4539&2.16&1.3&0.32&0.442&37&50$\pm$11&218&180$\pm$36&809&824$\pm$63\\
32&89573&4819&2.25&1.4&0.35&0.474&44&19$\pm$16&74&91$\pm$34&885& \\
33&89589&4690&2.13&1.5&-0.13&0.414&25&24$\pm$12&48&48$\pm$38&817&777$\pm$67\\
34&89590&5151&2.42&1.5&-0.23&0.419&37&53$\pm$14&115&120$\pm$36&798&854$\pm$62\\
35&89609&4898&2.29&1.2&-0.15&0.429&35&33$\pm$12&149&157$\pm$38&915&906$\pm$62\\
36&89614&4734&2.26&1.5&0.59&0.414&39&46$\pm$12&76& &784& \\
37&89640&4641&2.21&1.5&0.19&0.393&18&17$\pm$12&97&99$\pm$38&878&872$\pm$62\\
38&89645&4674&2.22&1.6&0.28&0.416&41&32$\pm$13&118&122$\pm$38&797&767$\pm$64\\
39&89667&4516&2.11&1.5&0.66&0.417&53&41$\pm$13&78&81$\pm$38&783&748$\pm$64\\
40&89669&4964&2.44&1.4&0.39&0.434&35&47$\pm$10&141&113$\pm$34&1029&1050$\pm$63\\
41&89687&4743&2.33&1.5&0.5&0.441&43&40$\pm$13&106&105$\pm$38&846&822$\pm$63\\
42&89702&4714&2.2&1.2&0.2&0.436&31&49$\pm$14&148&124$\pm$36&913&949$\pm$62\\
43&89731&4784&2.22&1.3&-0.12&0.421&33&31$\pm$13&129&137$\pm$35&825&803$\pm$72\\
44&89735&4579&2.07&1&0.11&0.45&36&25$\pm$13&111&116$\pm$38&976&944$\pm$64\\
45&89736&4483&2.15&1.4&0.47&0.431&30&50$\pm$12&255&210$\pm$36&837&850$\pm$63\\
46&89774&4635&2.13&1.5&0.22&0.44&39&38$\pm$10&150&145$\pm$37&820&834$\pm$62\\
47&89786&4705&2.31&1.1&-0.19&0.433&26&23$\pm$10&82&84$\pm$39&851&803$\pm$64\\
48&89802&4481&2.12&1.5&0.45&0.458&52&51$\pm$10&188&181$\pm$37&855&881$\pm$62\\
49&89832&4546&2.14&1.4&0.17&0.4&25&33$\pm$10&150&134$\pm$35&757&785$\pm$63\\
50&89843&5270&2.38&1.1&0.17&0.434&40&53$\pm$12&144&116$\pm$36&956&972$\pm$63\\
51&89848&4582&2.08&1.4&0.09&0.434&47&46$\pm$10&72& &931& \\
52&89871&4502&2.17&1.5&0.36&0.428&47&52$\pm$11&138&142$\pm$37&872&876$\pm$62\\
53&101453&4966&2.39&1.2&-0.1&0.433&35&34$\pm$11&160&167$\pm$38&928&920$\pm$62\\
54&101462&4807&2.25&1.3&0.07&0.433&24&37$\pm$13&167&147$\pm$36&883&904$\pm$63\\
55&101470&4567&2.2&1.4&0.62&0.415&5&24$\pm$17&22& &999& \\
56&101476&4838&2.34&1.5&-0.11&0.433&30&43$\pm$14&110& &802&858$\pm$63\\
57&101486&4760&2.15&1.3&-0.6&0.419&34&33$\pm$10&135&136$\pm$35&941&961$\pm$63\\
58&101505&4584&2.16&1.5&0.17&0.429&44&37$\pm$13&164&168$\pm$38&884&856$\pm$64\\
59&101540&4754&2.35&1.5&0.38&0.4&39&39$\pm$10&98&94$\pm$37&789&803$\pm$62\\
60&101606&4647&2.08&1.5&0.58&0.399&23&46$\pm$12&175&123$\pm$36&763&775$\pm$63\\
61&101638&4556&2.14&1.4&-0.37&0.408&28&27$\pm$10&145&140$\pm$37&650&654$\pm$63\\
62&101648&5025&2.33&1.6&-0.07&0.457&31&30$\pm$10&124&118$\pm$37&708&719$\pm$63\\
63&101650&4801&2.21&1.5&0.44&0.413&48&46$\pm$10&169&162$\pm$37&792&815$\pm$62\\
64&101655&5274&2.46&1.2&0.71&0.399&14&30$\pm$14&76&19$\pm$36&826&872$\pm$62\\
65&101662&4465&2.05&1.2&0.29&0.401&12&3$\pm$13&80&100$\pm$35&757&733$\pm$73\\
66&101674&4644&2.23&1.5&0.43&0.425&-5&19$\pm$14&122&83$\pm$36&808&840$\pm$62\\
67&101681&4763&2.23&1.5&0.36&0.433&22&35$\pm$14&82&46$\pm$36&654&701$\pm$62\\
68&101691&4857&2.23&1.6&0.21&0.412&30&30$\pm$10&137&134$\pm$37&649&658$\pm$62\\
69&101692&4449&1.99&1.5&0.25&0.41&21&20$\pm$12&145&146$\pm$38&707&702$\pm$62\\
70&101703&4931&2.2&1.5&0.16&0.404&41&31$\pm$13&99&103$\pm$38&691&662$\pm$64\\
71&234759&4458&2.06&1.5&0.35&0.507&6&31$\pm$13&204&169$\pm$36&1028&1055$\pm$63\\
72&234802&4288&1.97&0.8&0.23&0.488&55&64$\pm$10&229&202$\pm$35&1135&1154$\pm$63\\
73&234810&4521&2.12&1.2&-0.5&0.5&32&31$\pm$10&111&109$\pm$39&1054&1016$\pm$64\\
74&234811&4773&2.28&1.4&-0.39&0.498&24&23$\pm$11&81&78$\pm$39&1075&1029$\pm$64\\
75&234849&4556&2.19&1.2&0.2&0.467&57&56$\pm$12&188&190$\pm$38&961&956$\pm$62\\
76&234877&4538&2.14&0.8&-0.2&0.467&47&52$\pm$10&183&175$\pm$34&828&849$\pm$63\\
77&234898&4714&2.26&0.8&0.18&0.476&44&42$\pm$10&167&162$\pm$37&984&995$\pm$62\\
78&234905&5042&2.28&1.3&-0.12&0.478&30&31$\pm$9&143&138$\pm$37&836&847$\pm$63\\
79&234917&5050&2.39&1.5&-0.7&0.458&24&16$\pm$12&217&234$\pm$38&874&859$\pm$62\\
80&234926&4830&2.3&1.3&-0.68&0.445&36&33$\pm$10&140&136$\pm$37&934&928$\pm$63\\
81&234929&4893&2.35&0.8&-0.68&0.483&47&38$\pm$13&161&177$\pm$38&892&874$\pm$62\\
82&235017&4798&2.31&1.4&0&0.47&47&43$\pm$13&156&162$\pm$38&1024&1012$\pm$62\\
83&244618&4410&2.03&1.4&-0.17&0.463&42&40$\pm$10&161&157$\pm$37&899&905$\pm$63\\
84&244635&4650&2.24&1.5&-0.19&0.435&35&39$\pm$11&119& &855& \\
85&244642&4757&2.15&1.5&0.3&0.462&24&34$\pm$10&185&166$\pm$35&830&859$\pm$63\\
86&244660&4736&2.17&1.3&-0.28&0.43&44&39$\pm$13&139&149$\pm$38&890&878$\pm$62\\
87&244668&4627&2.17&1.5&0.51&0.406&44&62$\pm$11&167&119$\pm$36&868&880$\pm$63\\
88&244704&4916&2.22&1.3&-0.14&0.431&31&36$\pm$10&156&153$\pm$36&989&1012$\pm$63\\
89&244718&4724&2.29&1.4&0.12&0.417&30&31$\pm$9&177&171$\pm$37&960&974$\pm$63\\
90&244727&4641&2.27&1.3&0.35&0.394&36&29$\pm$13&107&109$\pm$38&823&792$\pm$64\\
91&244728&4491&2.03&1.5&0.3&0.463&39&50$\pm$14&140&112$\pm$36&940&987$\pm$62\\
92&244734&4900&2.26&1.5&0.52&0.452&52&44$\pm$11&140&147$\pm$39&811&754$\pm$64\\
93&244742&4735&2.23&1.5&-0.03&0.426&53&51$\pm$10&158&153$\pm$37&952&959$\pm$62\\
94&244757&4910&2.23&1.2&-0.53&0.419&29&46$\pm$14&107&126$\pm$36&857&908$\pm$62\\
95&244763&4775&2.19&1.5&0.14&0.396&33&33$\pm$10&143&136$\pm$37&806&822$\pm$63\\
96&244817&4610&2.11&1.5&0.57&0.439&40&54$\pm$10&149&109$\pm$35&866&884$\pm$63\\
97&244840&4605&2.11&1.5&0.45&0.412&21&27$\pm$11&156&157$\pm$38&916&914$\pm$62\\
98&256395&4650&2.21&1.6&0.41&0.418&27&27$\pm$13&143&142$\pm$38&896&892$\pm$62\\
99&256401&4650&2.29&1.5&0.54&0.415&47&32$\pm$13&28&59$\pm$34&931&906$\pm$73\\
100&256570&4843&2.19&1.5&0.56&0.42&32&44$\pm$16&97& &681& \\
101&256579&4737&2.26&1.5&0.36&0.422&17&17$\pm$12&115&114$\pm$38&699&695$\pm$62\\
102&256650&4598&2.19&1.5&0.44&0.412&26&30$\pm$12&124&123$\pm$38&674&672$\pm$62\\
103&267856&4605&2.2&1.2&0.04&0.419&60&61$\pm$11&137&141$\pm$37&771&769$\pm$62\\
104&267862&4686&2.23&1.1&-0.17&0.429&23&28$\pm$10&127&119$\pm$35&747&771$\pm$63\\
105&267909&4716&2.16&1.4&0.43&0.416&46&56$\pm$14&105&67$\pm$36&750&798$\pm$62\\
106&267912&4781&2.28&1.5&0.38&0.417&44&36$\pm$11&132&138$\pm$38&797&746$\pm$64\\
107&267985&4990&2.23&1.4&-0.14&0.411&16&12$\pm$11&85&85$\pm$39&660&606$\pm$64\\
108&268000&4568&2.2&1.5&0.19&0.416&31&35$\pm$12&29& &740& \\
109&268011&4696&2.24&1.7&0.4&0.419&36&36$\pm$10&126&120$\pm$37&751&767$\pm$62\\
110&268022&4968&2.26&1.5&0.22&0.385&24&27$\pm$11&128&131$\pm$37&712&714$\pm$62\\
111&268047&4880&2.34&1.3&0.09&0.437&31&28$\pm$10&96&99$\pm$38&749&691$\pm$64\\
112&268068&4837&2.31&1.7&0.3&0.416&39&31$\pm$11&144&151$\pm$39&767&698$\pm$64\\
113&385048&4686&2.24&1.2&-0.36&0.486&45&41$\pm$11&183&192$\pm$38&982&971$\pm$62\\
114&392992&4907&2.34&1.2&-0.22&0.484&40&44$\pm$10&163&156$\pm$34&977&998$\pm$63\\
115&393009&5012&2.32&1.5&-0.68&0.471&33&29$\pm$13&176&206$\pm$36&1054&1093$\pm$63\\
116&393012&4540&2.24&1.5&0.48&0.478&64&63$\pm$10&181&174$\pm$37&1006&1028$\pm$62\\
117&393024&4786&2.26&1.2&-1.13&0.468&35&48$\pm$12&112&103$\pm$38&1018&994$\pm$65\\
118&393026&4804&2.3&1.2&-0.5&0.497&41&42$\pm$10&168&171$\pm$35&1039&1061$\pm$63\\
119&393028&4457&2.13&1.4&0.36&0.461&18&14$\pm$10&103&109$\pm$39&1011&952$\pm$64\\
120&393030&4574&2.18&1.5&0.35&0.475&53&50$\pm$13&163&164$\pm$38&987&964$\pm$63\\
121&393033&4914&2.32&1&-0.5&0.453&43&37$\pm$11&213&224$\pm$38&976&964$\pm$62\\
122&393070&4903&2.3&1.6&0.47&0.459&41&37$\pm$10&134&141$\pm$38&1087&1023$\pm$64\\
123&393088&4575&2.21&1.5&0.22&0.466&20&41$\pm$13&164&137$\pm$36&979&1007$\pm$63\\
124&393091&4641&2.15&1.5&0.47&0.486&34&42$\pm$14&142& &904&944$\pm$62\\
125&393111&4546&2.21&1.5&0.23&0.464&31&45$\pm$14&157&130$\pm$36&1005&1048$\pm$62\\
126&393117&4931&2.26&0.9&-0.07&0.474&36&40$\pm$10&173&172$\pm$37&1015&1033$\pm$63\\
127&393124&4540&2.08&1.5&0.5&0.47&17&46$\pm$13&192&147$\pm$36&1025&1050$\pm$63\\
128&393125&4528&2.09&1.4&0.29&0.488&30&24$\pm$11&153&159$\pm$39&982&929$\pm$64\\
129&393136&4515&2.11&1.4&0.5&0.461&10&19$\pm$14&152& &1047&1085$\pm$63\\
130&393138&5073&2.34&1.4&-0.64&0.473&33&24$\pm$13&153&161$\pm$38&1020&999$\pm$64\\
131&393152&4757&2.23&1.3&-0.2&0.47&39&36$\pm$12&189&198$\pm$38&1019&1009$\pm$62\\
132&393162&4359&2.05&1.5&0.39&0.492&-15&7$\pm$18&161& &953& \\
133&393229&4868&2.32&1.3&-0.59&0.485&21&29$\pm$13&144&138$\pm$34&1021&1010$\pm$73\\
134&393319&4473&2.19&1.4&0.26&0.48&50&40$\pm$12&55&58$\pm$38&1045&981$\pm$64\\
135&393339&4972&2.41&1&-0.09&0.473&53&45$\pm$12&143&148$\pm$38&1007&968$\pm$64\\
136&402417&4655&2.2&1.2&-0.2&0.486&32&41$\pm$13&151&150$\pm$36&945&978$\pm$63\\
137&402421&4822&2.36&1.3&0.19&0.485&37&41$\pm$11&169&174$\pm$38&990&990$\pm$62\\
138&402432&4579&2.12&1.5&0.29&0.489&19&39$\pm$13&187&155$\pm$36&951&977$\pm$63\\
139&402433&4808&2.24&1.3&-0.19&0.461&36&43$\pm$12&149&144$\pm$36&989&1010$\pm$63\\
140&402435&4618&2.25&1.2&0&0.482&29&25$\pm$12&112&127$\pm$34&951&942$\pm$73\\
141&402440&4429&2.11&1.3&0.29&0.465&45&43$\pm$13&115&116$\pm$38&912&904$\pm$62\\
142&402449&4474&2.18&1.5&0.26&0.471&23&18$\pm$10&125&130$\pm$38&1042&978$\pm$64\\
143&402463&5260&2.43&1.2&-0.51&0.476&30&31$\pm$10&128&131$\pm$35&979&1001$\pm$63\\
144&402470&4270&1.97&1.2&0.16&0.457&27&35$\pm$16&111& &902& \\
145&402477&4545&2.19&1.5&0.51&0.432&37&49$\pm$10&196&175$\pm$35&827&859$\pm$63\\
146&402478&4983&2.35&1.2&-0.53&0.47&36&36$\pm$10&143&143$\pm$37&917&932$\pm$63\\
147&402482&4871&2.28&1.3&-0.57&0.478&34&30$\pm$10&118&120$\pm$39&910&875$\pm$64\\
148&402502&4818&2.19&1.2&-0.95&0.476&32&32$\pm$9&146&146$\pm$37&840&846$\pm$63\\
149&402516&4936&2.2&1.3&-0.65&0.46&27&22$\pm$10&130&133$\pm$39&791&761$\pm$64\\
150&402518&5034&2.32&1&-0.87&0.446&39&37$\pm$9&133&133$\pm$37&858&865$\pm$63\\
151&402550&4420&2.15&1.5&0.39&0.475&46&38$\pm$11&69&76$\pm$39&845&772$\pm$64\\
152&402560&4984&2.37&1.1&-1.01&0.441&35&35$\pm$9&146&145$\pm$37&825&826$\pm$63\\
153&402600&5158&2.4&1.3&-0.52&0.421&32&25$\pm$12&132&146$\pm$38&823&810$\pm$62\\
154&402621&4913&2.28&1.3&-0.56&0.45&37&35$\pm$13&142&164$\pm$36&767&804$\pm$63\\
155&402632&4937&2.32&1.5&0.55&0.439&34&30$\pm$10&130&137$\pm$39&635&572$\pm$64\\
156&402641&4775&2.18&1.5&0.49&0.458&23&36$\pm$10&120&90$\pm$34&782&803$\pm$63\\
157&412854&5191&2.4&1.4&-0.69&0.419&34&31$\pm$10&105&107$\pm$39&658&627$\pm$64\\
158&412889&4717&2.23&1.6&0.42&0.4&39&34$\pm$13&108&108$\pm$38&665&635$\pm$64\\
159&412894&5118&2.31&1.5&-0.45&0.427&25&30$\pm$12&99&97$\pm$34&724&715$\pm$73\\
160&412898&4871&2.22&1.8&0.07&0.415&22&18$\pm$10&126&130$\pm$38&627&570$\pm$64\\
161&412921&4617&2.13&1.5&0.47&0.413&69&79$\pm$14&99&60$\pm$36&601&647$\pm$62\\
162&412957&4920&2.23&1.5&-0.54&0.422&21&19$\pm$10&105&105$\pm$39&604&568$\pm$64\\
163&412979&4756&2.24&1.5&-0.58&0.424&36&32$\pm$10&134&133$\pm$37&652&643$\pm$62\\
164&413012&4869&2.29&1.2&-0.15&0.408&32&36$\pm$10&144&143$\pm$37&685&704$\pm$63\\
165&413043&4974&2.24&1.5&0.09&0.446&34&35$\pm$10&83& &699& \\
166&413060&4811&2.29&1.1&-0.37&0.422&29&31$\pm$10&136&136$\pm$37&724&741$\pm$63\\
167&413066&4674&2.11&1.5&0.44&0.419&33&33$\pm$10&193&186$\pm$37&684&709$\pm$63\\
168&423393&4702&2.28&1.5&0.41&0.408&24&25$\pm$10&150&143$\pm$37&696&718$\pm$63\\
169&423398&4351&2.05&1.3&0.42&0.426&35&47$\pm$10&182&156$\pm$34&646&672$\pm$63\\
170&423427&4818&2.22&1.5&0.33&0.479&41&30$\pm$12&126&132$\pm$38&688&646$\pm$64\\
171&423432&4945&2.36&1.4&-0.55&0.409&36&44$\pm$14&119&143$\pm$36&727&773$\pm$62\\
172&423434&4753&2.31&1.5&-0.15&0.502&28&25$\pm$12&145&153$\pm$38&759&749$\pm$62\\
173&423461&4743&2.24&1&-0.13&0.48&27&24$\pm$12&165&173$\pm$38&668&659$\pm$62\\
174&423485&4779&2.3&1.5&0.1&0.421&11&23$\pm$14&108&94$\pm$36&692&741$\pm$62\\
175&423486&5448&2.42&1.3&-0.39&0.483&28&32$\pm$13&112&121$\pm$36&695&723$\pm$63\\
176&423511&4472&2.03&1.2&-0.11&0.458&31&36$\pm$10&153&147$\pm$36&780&805$\pm$63\\
177&423516&4808&2.18&1.2&0.4&0.472&4&29$\pm$13&124&85$\pm$36&727&754$\pm$63\\
178&423530&4922&2.28&1&-0.11&0.415&51&49$\pm$10&152&148$\pm$37&807&811$\pm$62\\
179&423539&4676&2.11&1.5&0.54&0.418&36&50$\pm$10&144&104$\pm$35&774&791$\pm$63\\
180&423541&4491&2.18&1.4&0.2&0.447&39&34$\pm$10&116&120$\pm$39&710&649$\pm$64\\
181&423543&4975&2.4&1.4&-0.91&0.433&32&34$\pm$10&87&80$\pm$39&684&659$\pm$64\\
182&423551&4911&2.28&1.5&0.14&0.42&24&19$\pm$10&136&141$\pm$38&869&811$\pm$64\\
183&423569&4598&2.18&1.5&0.45&0.442&42&52$\pm$10&183&180$\pm$37&688&711$\pm$63\\
184&423573&5164&2.39&1.1&-0.59&0.421&25&24$\pm$10&119&120$\pm$37&764&780$\pm$63\\
185&423604&4422&2.07&1.3&0.55&0.422&21&21$\pm$13&96&93$\pm$38&736&731$\pm$62\\
186&423876&4671&2.29&1.5&0.02&0.469&36&42$\pm$12&105& &618& \\
187&545299&4483&2.16&1&0.17&0.486&52&48$\pm$13&144&147$\pm$38&1111&1093$\pm$62\\
188&545326&4578&2.16&1.1&0.2&0.524&8&20$\pm$14&102& &1260&1307$\pm$63\\
189&545421&5012&2.39&1.2&-0.22&0.492&45&49$\pm$10&170&162$\pm$34&1047&1068$\pm$63\\
190&545444&4407&2.07&1.5&0.48&0.485&23&15$\pm$11&63&69$\pm$39&991&936$\pm$64\\
191&545498&4806&2.26&1.5&-0.5&0.484&31&31$\pm$10&159&160$\pm$37&995&1012$\pm$63\\
192&545519&4917&2.41&1.5&-0.31&0.483&58&60$\pm$10&169& &1173& \\
193&545749&4556&2.25&1.4&0.04&0.485&44&42$\pm$12&155&160$\pm$38&1003&995$\pm$62\\
194&554729&4752&2.28&1.2&-0.83&0.474&41&38$\pm$10&162&190$\pm$36&997&1025$\pm$63\\
195&554738&4701&2.32&1.2&0.13&0.511&40&38$\pm$10&148&142$\pm$37&1007&1020$\pm$62\\
196&554754&5056&2.3&1.4&-0.52&0.485&41&35$\pm$11&191&201$\pm$38&949&937$\pm$62\\
197&554762&4625&2.29&1.4&0.13&0.484&44&45$\pm$11&155&157$\pm$37&959&960$\pm$62\\
198&554807&4941&2.4&1.4&-0.04&0.484&32&28$\pm$13&92&104$\pm$34&887&867$\pm$73\\
199&554820&4824&2.37&1.5&0.24&0.483&33&22$\pm$12&84&99$\pm$34&815& \\
200&554835&4622&2.17&1.3&0.48&0.455&13&30$\pm$17&105& &882& \\
201&554849&4581&2.19&1.5&0.46&0.47&43&61$\pm$12&225&180$\pm$36&861&874$\pm$63\\
202&554860&4877&2.24&1.3&-0.32&0.477&38&32$\pm$13&147&157$\pm$38&814&802$\pm$62\\
203&554876&4949&2.38&1.5&-0.03&0.447&29&24$\pm$10&131&135$\pm$38&748&696$\pm$64\\
204&554901&4638&2.11&1.2&-0.67&0.45&31&34$\pm$12&99&93$\pm$38&714&677$\pm$64\\
205&564834&5356&2.41&1.3&-0.36&0.424&34&32$\pm$10&134&130$\pm$37&772&773$\pm$63\\
206&564840&4713&2.23&1.5&0.33&0.435&23&8$\pm$12&37&54$\pm$34&734& \\
207&564857&4912&2.28&1.4&0.01&0.448&21&21$\pm$10&122&116$\pm$37&695&708$\pm$63\\
208&564949&4619&2.14&1.5&0.4&0.433&23&45$\pm$13&192&153$\pm$36&765&787$\pm$63\\
209&564992&5269&2.37&1.5&-0.61&0.432&30&29$\pm$10&142&143$\pm$37&685&700$\pm$63\\
210&565019&4801&2.29&1.5&0.38&0.427&46&36$\pm$12&121&127$\pm$38&656&611$\pm$64\\
211&565025&4729&2.19&1.4&-0.58&0.426&32&29$\pm$10&118&119$\pm$38&629&594$\pm$64\\
212&565056&5067&2.39&1.6&0.63&0.43&60&68$\pm$14&56&2$\pm$37&717&766$\pm$62\\
213&565094&5049&2.32&1.5&0.14&0.515&26&20$\pm$13&119&131$\pm$35&718&687$\pm$71\\
214&575390&4639&2.1&1.5&0.18&0.494&40&47$\pm$10&147&144$\pm$37&692&714$\pm$63\\
215&575421&4685&2.19&1.2&-0.08&0.477&31&30$\pm$10&162&157$\pm$37&658&669$\pm$63\\
216&575448&4583&2.19&1.3&0.58&0.467&39&27$\pm$13&25&42$\pm$36&675&631$\pm$69\\
217&575485&4634&2.28&1.1&-0.09&0.483&32&35$\pm$9&189&186$\pm$37&707&721$\pm$63\\
218&575542&4636&2.21&1.5&0.58&0.425&24&24$\pm$13&124&120$\pm$38&703&699$\pm$62\\
219&575585&4769&2.19&1.2&-0.17&0.441&35&27$\pm$12&104&109$\pm$38&671&636$\pm$64\\

 \end{longtable}}

\begin{appendix} 
\section{Error estimates}

Estimating uncertainties on the DIB equivalent widths is not straightforward here, as there are different sources of errors. The error linked to measurement uncertainties can in principle be classically estimated.
 For the 219 spectra of this study, the signal-to-noise ratio (S/N) varies between from 30 and 77 per pixel. It can be immediately derived that for a broad DIB covering about 6 \AA\ , i.e. defined over more than 100 pixels, and of about 10-15 \% depth, which corresponds to the 6284 \AA\ one, the relative error due to the noise only should be very small. For the shallower and narrower DIBs this is not the case.
The other and major source of uncertainties comes from the use of synthetic spectra, that have not been adjusted individually for all targets. As shown by figures 6 to 9 , those uncertainties may become of the order of the DIB itself  depending on the star radial velocity, for the two small DIBs.

\begin{figure*}
   \centering
   \includegraphics[width=9cm]{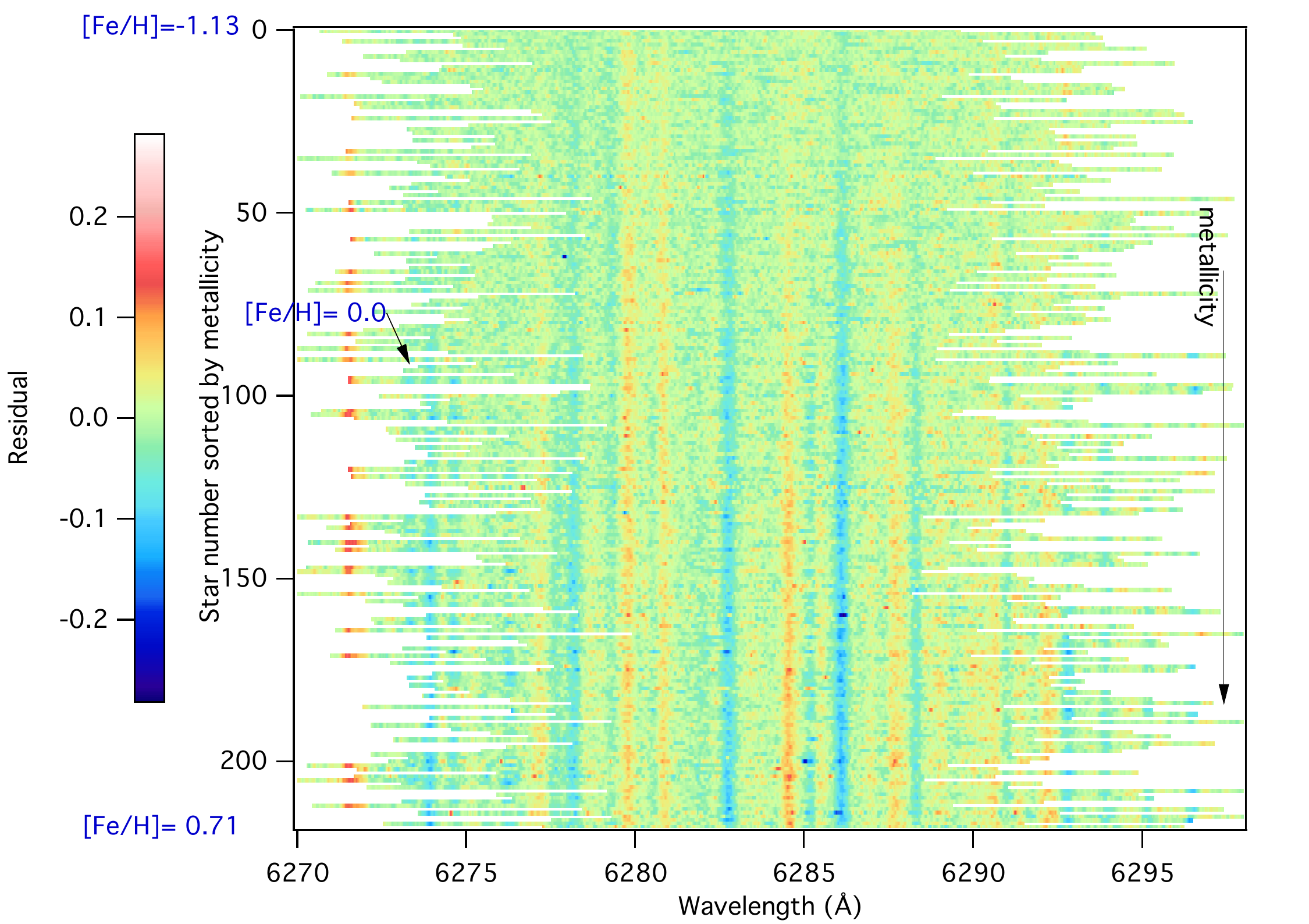}
     \caption{Residuals as a function of wavelength and metallicity  for the DIB 6283.8 \AA\ . They form the basis for the EW correction and the error estimate.}
         \label{residual_6280A}
   \end{figure*}

\begin{figure*}
   \centering
   \includegraphics[width=9cm]{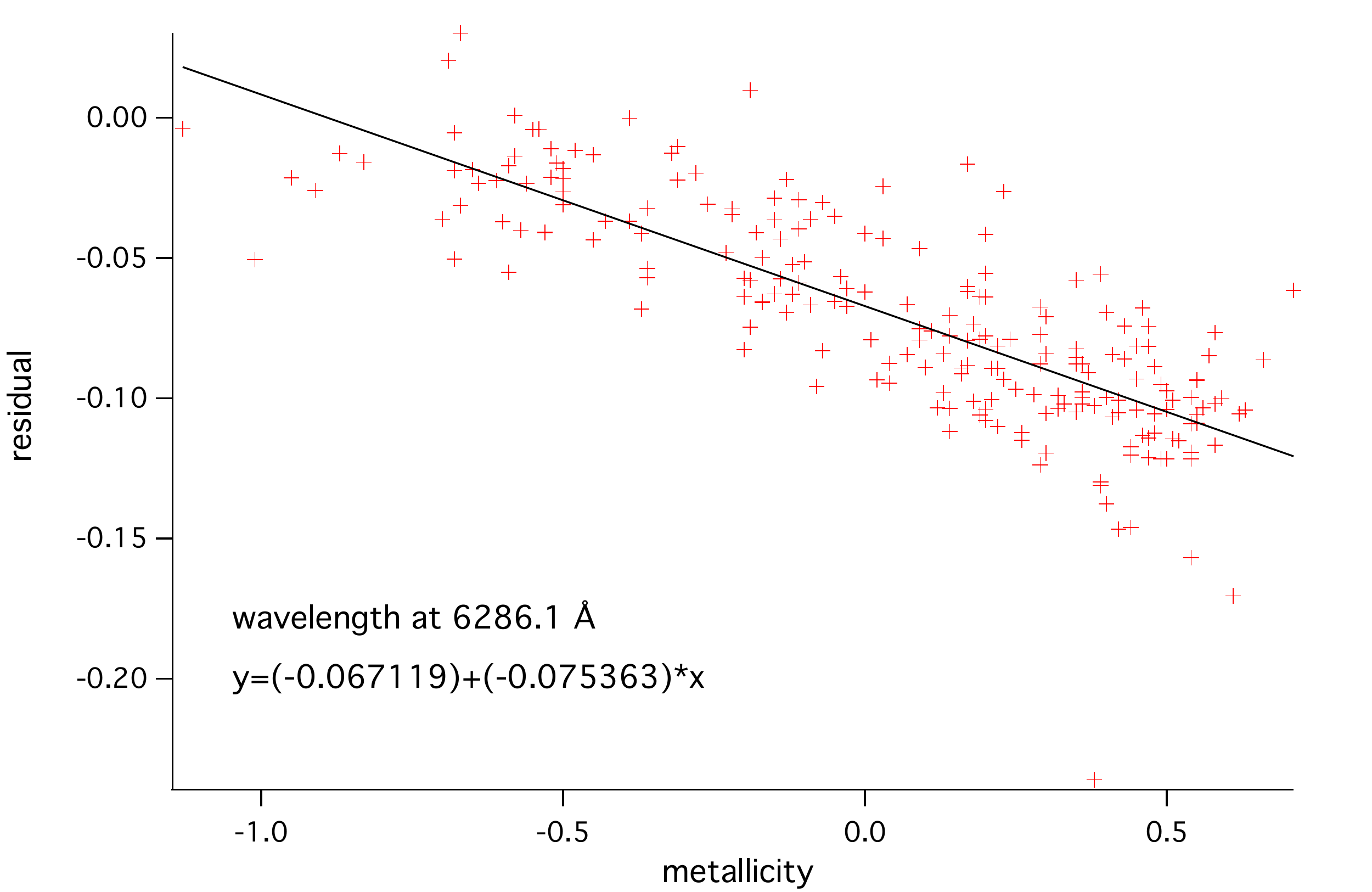}
     \caption{Example of residuals vs -metallicity relationships: here the residuals depend on the metallicity and are negative, which corresponds to an under-predicted stellar absorption line.}
         \label{resi_FeH_fitting}
   \end{figure*}

\begin{figure*}
   \centering
   \includegraphics[width=9cm]{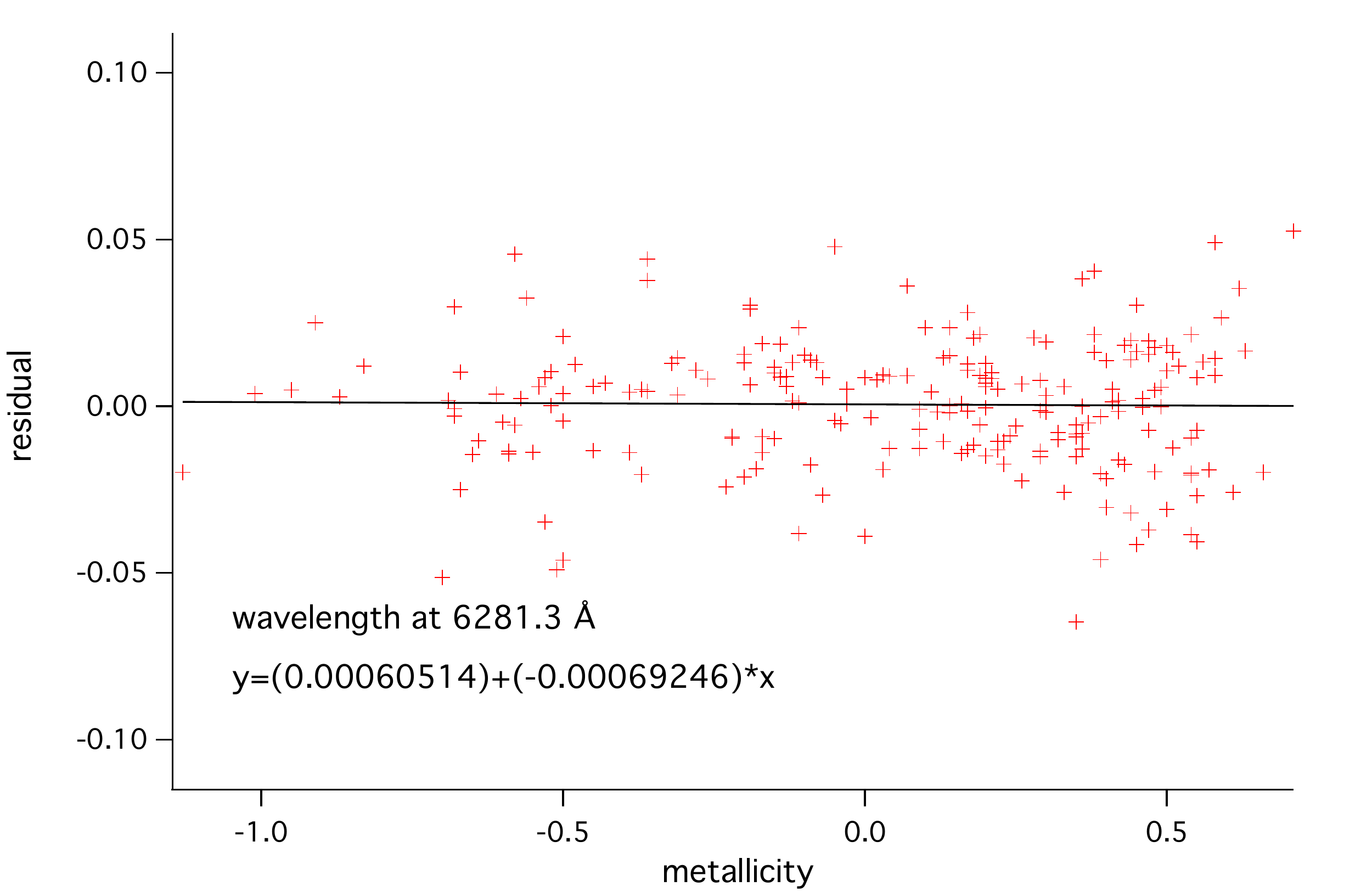}
     \caption{Example of residuals vs -metallicity relationships: here the residuals do not depend on the metallicity and are negligible.}
         \label{resi_FeH_fitting_no_relation}
   \end{figure*}

\begin{figure*}
   \includegraphics[width=9cm]{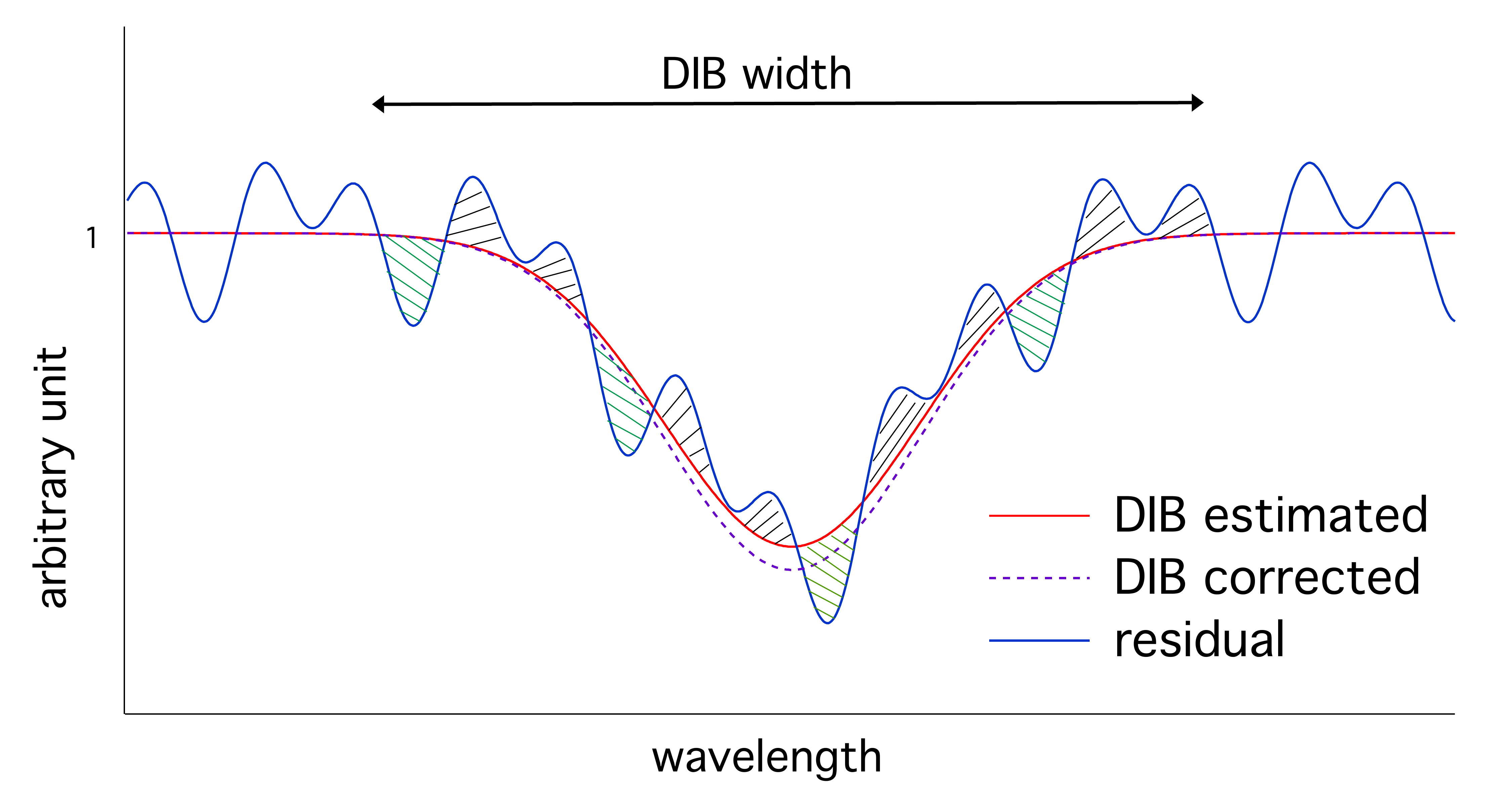}
     \caption{Schematic illustration of the correction applied to the EW measurements. The red curve represents the DIB profile as it comes out from the fitting phase. Hatched surfaces show the offsets applied at each wavelength that are computed for each star, as a function of its metallicity. The dashed line illustrates the corrected DIB.  
              }
         \label{dib_example}
   \end{figure*}

We have devised a method for error estimates that has the advantage of providing a corrective term for the DIB equivalent width  in addition to the errors. For each target, the $(data - model)$ residuals for the best adjustment are shifted into the stellar frame. Those residuals are then sorted by metallicity. The corresponding 2D residual plot is  shown in Figure \ref{residual_6280A} for the 6284 \AA\ DIB. The x-axis is the wavelength in the star frame, and the Y axis is the metallicity. The color scale represents the value of the residuals. Obviously  there are some features, some quite strong, at specific wavelengths and they are also obviously related to the metallicity. They correspond to stellar lines that are under- or over-predicted by the synthetic spectra. A large number of lines are simply unidentified. The next step is to use this map in the orthogonal direction, i.e., to extract the residual as a function of the metallicity, for each wavelength in the star frame. Examples of the obtained extracted series are shown in Figures \ref{resi_FeH_fitting} and  \ref{resi_FeH_fitting_no_relation}. Each of those metallicity-residual series of points is then fitted to a linear relationship, as shown in the two plots. For the wavelength corresponding to Fig. \ref{resi_FeH_fitting}, the residual is negative and its absolute value is strongly increasing with the metallicity. This means that at this location is a stellar line that is under-predicted and that the larger the metallicity, the larger the data-model discrepancy. 
At variance with this case, at the wavelength corresponding to Fig. \ref{resi_FeH_fitting_no_relation}, the residual is in average zero and not related to the metallicity. This means that there is no stellar line at this location, or that the model does predict the line adequately. 
 We then use both the fit coefficients (at each wavelength) to perform a correction to the previously computed DIB profile, and the standard deviation around the mean relationship (again at each wavelength) to estimate the random (or quasi-random) remaining errors. 
More precisely, for each star we consider the wavelength interval that contains the entire DIB, and compute for each wavelength, based on the linear relationships coefficients, the most-probable residual as a function of the star metallicity. We apply this correction over the whole DIB interval (as shown in Figure \ref{dib_example}) and recalculate the DIB equivalent width. Table \ref{DIB_index} contains both the initial and corrected values for the EW for the three DIBS and all targets. We found that the recalculated EWs provide significantly improved DIB-DIB correlations, which demonstrates that this method provides a partial, but valid correction.

The distribution of data points around the mean residual-metallicity relationship provides an estimate of the combination of unpredictable uncertainties linked to the stellar lines and the actual noise). We then used the standard deviation measured for each relationship  to computed the standard deviation at each wavelength, then propagated the errors on the whole DIB interval. In the case of the two narrow DIBs, this estimated uncertainty may be in some cases smaller than the actual errors, however considering those devaitions as systematic and not random leads to errors that are irrealistically large.

%********************************

\end{appendix}

\end{document}